\documentclass[]{agujournal2019}
\usepackage{url} 
\usepackage{amssymb} 
\usepackage{xcolor}
\usepackage{lineno}
\usepackage{subcaption}
\usepackage{afterpage}
\usepackage[inline]{trackchanges} 
\usepackage{soul}
\usepackage[english]{babel}
\usepackage{setspace}
\usepackage{microtype}
\usepackage{multirow}
\usepackage{bm}
\usepackage{natbib}
\usepackage[unicode=true, colorlinks=true, linkcolor=blue, urlcolor=blue, citecolor=blue, anchorcolor=blue]{hyperref}
\usepackage{xcolor} 
\usepackage{soul}   
\usepackage{booktabs}
\usepackage{amsmath}
\usepackage{array}
\usepackage{graphicx}	
\usepackage{orcidlink}

\draftfalse

\journalname{Space Weather}

\usepackage{ragged2e}  

\begin{document}
	\justify

\title{Operational Solar Flare Forecasting System Using an Explainable Large Language Model}

\authors{Xuebao Li\orcidlink{0000-0003-0397-4372}\affil{1,2}\thanks{Xuebao Li, Yongshang Lv, and Jinfang Wei contributed equally to this work.}, Yongshang Lv\orcidlink{0009-0004-9110-0425}\affil{1,2}, Jinfang Wei\orcidlink{0009-0001-3404-3186}\affil{3,1}, Yanfang Zheng\orcidlink{0000-0003-0229-3989}\affil{1,}\thanks{Corresponding author. Email: \href{mailto:zyf062856@163.com}{zyf062856@163.com}}, Ting Li\orcidlink{0000-0001-6655-1743}\affil{2,4,5}, Rui Wang\orcidlink{0000-0001-5205-1713}\affil{2}, Zixian Wu\orcidlink{0009-0005-2367-9647}\affil{1}, Hongwei Ye\orcidlink{0009-0009-9722-5794}\affil{1}, Pengchao Yan\orcidlink{0000-0002-3667-3587}\affil{1}, Zamri Zainal Abidin\orcidlink{0000-0002-7149-0997}\affil{6,7}, Noraisyah Mohamed Shah\orcidlink{0000-0002-1320-8711}\affil{8}, Changtian Xiang\orcidlink{0009-0009-8072-3519}\affil{1}, Shunhuang Zhang\orcidlink{0009-0003-0325-1366}\affil{1}, Xiaojia Ji\orcidlink{0009-0004-6402-7509}\affil{1}, Xusheng Huang\orcidlink{0009-0008-1279-1870}\affil{1}, Xiaotian Wang\orcidlink{0009-0004-1408-8055}\affil{1}, Honglei Jin\orcidlink{0009-0006-5294-3571}\affil{1}
}

\affiliation{1}{School of Computer Science, Jiangsu University of Science and Technology, 212100 Zhenjiang, China}
\affiliation{2}{State Key Laboratory of Solar Activity and Space Weather, National Space Science Center, Chinese Academy of Sciences, Beijing 100190, China}
\affiliation{3}{School of Software, Southeast University, Nanjing, China}
\affiliation{4}{National Astronomical Observatories, Chinese Academy of Science, Beijing 100101, China}
\affiliation{5}{School of Astronomy and Space Sciences, University of Chinese Academy of Sciences, Beijing, China}
\affiliation{6}{Radio Cosmology Lab, Centre for Astronomy and Astrophysics Research, Department of Physics, Faculty of Science, Universiti Malaya, 50603 Kuala Lumpur, Malaysia}
\affiliation{7}{National Centre for Particle Physics, Universiti Malaya, 50603 Kuala Lumpur, Malaysia}
\affiliation{8}{Department of Electrical Engineering, Faculty of Engineering, University of Malaya}

\begin{keypoints}
	
	\item{By employing BERT as a universal computation engine, LLMFlareNet pioneers solar flare forecasting with superior performance.}
	\item{Using SHAP for explainability analysis, we reveal a strong correlation between R\_VALUE and the flare predictions generated by LLMFlareNet.}
	\item{A new "daily mode" compares operational forecasting systems, showing LLMFlareNet-based system surpasses NASA/CCMC and SolarFlareNet.}
	
\end{keypoints}

\begin{abstract}

This study focuses on forecasting major ($\geq$M-class) solar flares that can severely impact the near-Earth environment. We construct two types of datasets using the Space Weather HMI Active Region Patches (SHARP), and develop a flare prediction network based on large language model (LLMFlareNet). We apply SHapley Additive exPlanations (SHAP) to explain the model predictions. We develop an operational forecasting system based on the LLMFlareNet model.
We adopt a daily mode for performance comparison across various operational forecasting systems under identical active region (AR) number and prediction date, using daily operational observational data. The main results are as follows. (1) Through ablation experiments and comparison with baseline models, LLMFlareNet achieves the best TSS scores of 0.720$\pm$0.040 on the ten cross-validation (CV) dataset with mixed ARs. (2) By both global and local SHAP analyses, we identify that R\_VALUE is the most influential physical feature for the prediction of LLMFlareNet, aligning with flare magnetic reconnection theory. (3) In daily mode, LLMFlareNet achieves TSS scores of 0.680/0.571 (0.689/0.661, respectively) on the dataset with single/mixed ARs, markedly outperforming NASA/CCMC (SolarFlareNet, respectively). This work introduces the first application of a large language model as a universal computation engine with explainability method in this domain, and presents the first comparison between operational flare forecasting systems in daily mode. The proposed LLMFlareNet-based system demonstrates substantial improvements over existing systems.
\end{abstract}
\textbf{Keywords:} Solar activity (1475) --- Solar flares (1496) --- Solar active region magnetic fields (1975) --- Astronomy data analysis (1858)

\section*{Plain Language Summary}

This study advances solar flare forecasting with the LLMFlareNet model achieving superior performance in predicting $\geq$M-class flares within 24 hr. Through SHAP explainable analysis, we identify the strong correlation between the prediction of LLMFlareNet and R\_VALUE. The operational forecasting system based on LLMFlareNet is significantly superior to existing systems such as NASA/CCMC and SolarFlareNet.

\section{Introduction} \label{sec:intro}

Solar flares, especially major ($\geq$M-class) flares, are a violent explosive phenomenon in solar activity, where the magnetic disturbance generated can release a large amount of magnetic energy, primarily in the form of electromagnetic radiation and high-energy particles released into space \citep{priest2002magnetic}. The energy released by solar flares not only has a profound impact on the space environment of the solar system but also affects human production activities. Especially during intense flare events, the released high-energy particles and electromagnetic radiation can significantly disturb the near-Earth space environment, causing communication interruptions and affecting the accuracy of navigation systems, thereby endangering the safety of the aerospace field \citep{baker2004effects,schou2012design}. Timely and accurate forecasting of solar flares can provide valuable time for implementing countermeasures, thereby minimizing losses to the greatest extent. Therefore, developing an efficient and accurate operational flare forecasting system has significant practical application value.

Deep learning, as an advanced branch of machine learning, has been widely utilized in flare prediction (\citealt{RN6,park2018application,liu2019predicting,li2020predicting}). Deep learning has the ability to learn more complex and abstract features from raw observational data through multiple layers of nonlinear transformations, thereby enhancing the accuracy and efficiency of models. 
\citet{RN6} was the first to successfully develop a convolutional neural network (CNN) model for solar flare forecasting. Subsequently, researchers have explored various deep learning architectures for this task, such as applying CNNs to perform binary flare prediction (e.g.,~\citealt{li2020predicting, park2018application}), or incorporating long short-term memory (LSTM) networks to enhance temporal modeling capability (e.g., \citealt{liu2019predicting, guastavino2022implementation, tang2021solar}).
To better capture temporal dependencies, long-range relationships, and multidimensional information in the data, the transformer architecture network \citep{vaswani2017attention} of deep learning has been introduced into solar flare forecasting. Compared to traditional deep learning such as CNN and LSTM, which can only capture local contextual relationships, the transformer network can effectively capture the relationships between any two positions in the time series through the attention mechanism, thereby enabling it to learn global contextual information.
\citet{kaneda2022flare} developed a hybrid CNN and transformer model that utilizes full-disk magnetograms and sunspot region features from the Helioseismic and Magnetic Imager onboard the Solar Dynamics Observatory (SDO/HMI; \citealt{pesnell2012solar,schou2012design}) to predict $\geq$M-class flares. \citet{abduallah2023operational} constructed a transformer model that used sequential physical parameters to predict $\geq$M5.0, $\geq$M, and $\geq$C-class flares, respectively. \citet{grim2024solar} proposed multiscale vision transformer (MViT) model that utilized sequences of solar magnetic field images to predict $\geq$M-class solar flares. Additionally, other researchers have also employed transformer-based models for solar flare prediction tasks (e.g., \citealt{li2024prediction2,alshammari2024transformer, pelkum2024forecasting}).

With the continuous enhancement of pre-trained parameters and the increasing volume of training data, transformer-based Large Language Models (LLMs) have been developed and applied, demonstrating stronger generalization capabilities in complex tasks \citep{brown2020language}. The knowledge accumulated by LLMs across different tasks and domains can be fully utilized in specific tasks through transfer learning, improving the adaptability of the model in new domains. Since solar flares mainly originate from the dynamic evolution of magnetic fields of active regions (ARs), solar flare forecasting can essentially be regarded as a typical time-series classification problem.
\citet{lee2020time} pointed out that most magnetic parameter time series of ARs analyzed in their study are non-stationary, with different physical features exhibiting distinct persistent trends. This characteristic suggests the need for operational solar flare forecasting models to be capable of capturing long-range and complex temporal dependencies, while also maintaining strong adaptability to continuously evolving data distributions. Traditional models, such as LSTMs, are limited by memory decay when handling non-stationary sequences with long-range dependencies. While standard Transformers improve the modeling of long-range dependencies via self-attention mechanisms, they are typically trained on domain-specific, limited datasets. As a result, the temporal patterns learned by Transformers struggle to generalize under distribution shifts on non-stationary time series data \citep{Liu2022Nonstationary}. In contrast, LLMs can perform non-data-dependent operations through pre-training on massive and diverse datasets \citep{zhou2023one}. Such pre-training endows LLMs with a generic function that may provide a mechanism for modeling long-range dependencies and complex nonlinear interactions among features in non-stationary solar physics time series.
Recent studies have successfully applied LLMs to time-series tasks. \citet{zhang2024large} summarized five approaches for leveraging LLMs in time-series applications, including direct prompting of LLMs, time series quantization, aligning techniques, utilization of the vision modality as a bridging mechanism, and the combination of LLMs with tools. For instance, \citet{li2025deep} employed a prompting approach using LLMs to perform automatic classification of variable star light curves. In contrast, \citet{lu2022frozen} proposed a new application framework, Frozen Pretrained Transformer (FPT), pointing out that a pre-trained LLM can be used as a universal computation engine for non-language downstream tasks. \citet{zhou2023one} applied LLMs to time-series forecasting via the FPT framework. Their theoretical and experimental analysis reveals that the self-attention mechanism in LLMs inherently resembles principal component analysis (PCA). This finding supports the use of the LLM as a universal computation engine for time-series tasks. However, to the best of our knowledge, no prior work has introduced the LLM as a universal computation engine into the domain of solar flare forecasting.

Although deep learning models have been widely applied to solar flare forecasting, their internal decision mechanisms remain complex and opaque, raising concerns about their trustworthiness in scientific research and operational applications. Therefore, incorporating explainable artificial intelligence (XAI) techniques to reveal which features the model focuses on and how they influence its decisions has become a key step toward building reliable deep learning models. Shapley Additive exPlanations (SHAP;~\citealt{lundberg2017unified}) is one of the most widely adopted explainability techniques. By leveraging cooperative game theory to compute the contribution of each input feature to the model prediction, SHAP provides comprehensive analysis of model explainability at both the global and local level. In recent years, several studies have applied SHAP to the explainability analysis of space weather forecasting models, thereby improving the credibility of these models (e.g., \citealt{ye2024evaluating,gazula2024interpretable,rawashdeh2025explainable}).

In the field of solar flare prediction, while the improvement of model performance is crucial, the construction of operational forecasting system is equally significant and cannot be overlooked. An efficient, timely, and accurate forecasting system can not only quickly respond to solar flare events but also accurately transmit prediction results to relevant departments and users, thereby providing timely decision support for addressing potential space weather impacts. \citet{nishizuka2021operational} used DNNs to develop an operational solar flare forecasting system. \citet{abduallah2023operational} proposed a transformer-based SolarFlareNet model and developed an operational forecasting system to predict $\geq$M5.0-class, $\geq$M-class, and $\geq$C-class flares (\url{https://nature.njit.edu/solardb/index.html}). \citet{yan2024real}  developed an operational full-disk solar flare forecasting system based on five deep learning models to forecast the occurrence of $\geq$C-class and $\geq$M-class flares. Additionally, the Community Coordinated Modeling Center of the National Aeronautics and Space Administration (NASA/CCMC; \citealt{hesse2001community}) integrates multiple models to provide operational solar flare prediction results (\url{https://ccmc.gsfc.nasa.gov/scoreboards/flare/}). Currently, some forecasting systems have been applied to the field of solar flare prediction, but no scholars have yet conducted a comparative analysis of the performance of operational AR flare forecasting systems.

In this study, we establish two types of datasets for $\geq$M-class
flare prediction. The first type of dataset is the ten cross-validation (CV) datasets for model training, validation, and testing in Section \ref{subsec:Ten CV data}. The second type of dataset is comparison datasets used for model testing in Section \ref{subsec:Compa data}. We are the first to introduce an LLM as a universal computation engine for solar flare forecasting and propose a flare prediction network based on an LLM (LLMFlareNet) for predicting $\geq$M-class flares within 24 hr. To verify the effectiveness of using an LLM as a universal computation engine, we conduct systematic ablation experiments. In addition, we build the LSTM and the Neural Network (NN) model as baseline models and compare their performance with that of LLMFlareNet. Additionally, we apply the SHAP method to perform explainability analysis on LLMFlareNet, quantifying the contribution of each physical feature to the model prediction. Based on the LLMFlareNet, we develop an operational forecasting system for AR solar flares within 24 hr, and compare the prediction performance of our system ​​with​​ that of other operational forecasting systems (e.g., NASA/CCMC, SolarFlareNet) in daily mode. The rest of this paper is organized as follows. The data is described in Section \ref{sec:Data}, and the method is introduced in Section \ref{sec:Meth}. Results are given in Section \ref{sec:Res}, and finally, conclusions and discussions are provided by Section \ref{sec:conclu}.

\section{Data} \label{sec:Data}

\subsection{Ten CV datasets} \label{subsec:Ten CV data}
The HMI onboard the SDO satellite has been delivering high-resolution photospheric magnetic field data since 2010. The Space Weather HMI Active Region Patches (SHARP) data offers line-of-sight and vector magnetograms of ARs along with associated physical parameters \citep{bobra2014helioseismic}. We collect four classes of SHARP data from May 1, 2010, to February 13, 2022. The labeling process is identical to that of \citet{zheng2023comparative}. Firstly, we continuously observe the behavior of a specific AR for 24 hr. If no flare with an intensity exceeding C1.0 class occurs within this period, the AR and its associated magnetic field image samples are labeled as N-class (intensity less than C1.0). Secondly, if a C-class, M-class, or X-class flare occurs within the observed 24 hr, the AR is annotated with the corresponding category based on the level of flare eruption. It is noteworthy that if the same AR produces flares on different days or multiple times within a single day, we only retain the AR data of the highest flare level. Thirdly, we adopt a four-level AR classification scheme based on the maximum GOES-level flare an AR ever yields, consistent with \citet{zheng2023comparative} and \citet{li2020predicting}. In addition, although the data includes C-class, M-class, and X-class flares, flares of $\leq$C-class generally do not cause significant space weather impacts. Therefore, our study focuses on forecasting major ($\geq$M-class) flares to meet operational requirements. We take magnetograms every 36 minutes, resulting in a final total of 40 magnetogram samples for each AR.

\begin{table*}
	\centering
	\caption{Brief description and formula of ten magnetic field parameters from SHARP}. \label{tab1}%
	\makebox[\textwidth][c]{%
		\begin{tabular}{l l l }
			\hline
			Keyword  & Description                                                                                   & Formula                                             \\ \hline
			TOTUSJZ   & Total unsigned vertical current            & $J_{Z_{total}}=\sum \mid J_{Z}\mid dA$ \\
			TOTUSJH   & Total unsigned current helicity                                                               & $H_{C_{total}}\propto \sum \mid B_{Z}\cdot J_{Z} \mid$\\
			TOTPOT    & Total photospheric magnetic free energy density     & $\rho _{tot}\propto \sum \left ( B^{Obs} -B^{Pot}\right )^{2}dA $                                                 \\
			ABSNJZH   &  Absolute value of the net current helicity   & $H_{C_{abs}}\propto \mid \sum B_{Z} \cdot J_{Z}\mid$                                                    \\
			SAVNCPP   &  Sum of the modulus of the net current per polarity & $J_{Z_{sum}}\propto \mid \sum_{}^{B_{Z}^{+}}J_{Z}dA \mid +\mid \sum_{}^{B_{Z}^{-}}J_{Z}dA \mid$                                                    \\
			USFLUX    & Total unsigned flux                                                                           & $\varphi =\sum \mid B_{Z} \mid dA$                                                    \\
			AREA\_ACR &  Area of strong field pixels in the active region  &  $Area=\sum Pixels$                                                   \\
			MEANPOT   &  Mean photospheric magnetic free energy        & $\bar{\rho }\propto \frac{1}{N}\sum \left ( B^{Obs}-B^{Pot} \right )^{2}$                                                    \\
			R\_VALUE  & Sum of flux near polarity inversion line       & $\varphi =\sum \mid B_{LoS}\mid dA \ within \ R \ m ask$                                                     \\
			SHRGT45   & Fraction of area with shear \textgreater{}45°     & $Area\ with\ shear > 45^{\circ} / total\_area$                                                  \\ \hline
		\end{tabular}
	}
\end{table*}

We use 10 magnetic field parameters, such as TOTUSJH, TOPOT, TOTUSJZ, ABSNJZH, SAVNCPP, USFLUX, AREA\_ACR, MEANPOT, R\_VALUE, and SHRGT45 in our work. Table \ref{tab1} illustrates a brief description and formula of the ten magnetic field parameters from SHARP. In the formula of R\_VALUE, the $R \ m ask$ identifies the areas within about 15 Mm of high-gradient strong-field polarity-separation lines, as described by \citet{schrijver2007characteristic}. We extract 10 physical feature parameters from all magnetograms involved in our work and create ten CV datasets based on the AR segmentation method. These constitute the first type of dataset used for model training, validation, and testing. Additionally, we normalize the ten CV datasets using the z-score method \citep{zscore}, which applies the mean and standard deviation calculated from the entire CV set.
The ten CV datasets are identical to those created by \citet{zheng2023comparative}, with the testing dataset denoted as the testing dataset with mixed ARs. Furthermore, to investigate whether the physical parameters in magnetograms containing multiple ARs affect model performance, we retain physical parameters in magnetograms containing only single AR from testing dataset. Then, we create 10 filtered CV testing datasets, referred to as the testing dataset with single AR. Table \ref{tab2} presents the number distribution  of samples and ARs in the testing dataset with mixed/single ARs. The number distribution of the other nine testing datasets with mixed ARs is identical to the one displayed in Table \ref{tab2}.

\begin{table*}
	\raggedright
	\renewcommand{\arraystretch}{1.2}
	\centering
	\caption{The number distribution of samples and ARs in the testing dataset with mixed/single ARs.}\label{tab2}%
	\makebox[\textwidth][c]{%
		\begin{tabular}{ccccc}
			\hline
			Testing   datasets                & N(Sample/AR) & C(Sample/AR) & M(Sample/AR) & X(Sample/AR) \\ \hline
			Testing dataset with mixed ARs  & 3200/80      & 2400/60      & 1160/29      & 240/6        \\
			Testing dataset0 with single AR & 2320/58      & 1320/33      & 640/16       & 160/4        \\
			Testing dataset1 with single AR & 2520/63      & 1480/37      & 640/16       & 120/3        \\
			Testing dataset2 with single AR & 2360/59      & 1200/30      & 560/14       & 160/4        \\
			Testing dataset3 with single AR & 2120/53      & 1160/29      & 680/17       & 200/5        \\
			Testing dataset4 with single AR & 2240/56      & 1200/30      & 520/13       & 160/4        \\
			Testing dataset5 with single AR & 2000/50      & 1200/30      & 680/17       & 120/3        \\
			Testing dataset6 with single AR & 2440/61      & 1440/36      & 640/16       & 160/4        \\
			Testing dataset7 with single AR & 2320/58      & 1320/33      & 760/19       & 80/2         \\
			Testing dataset8 with   single AR & 2400/60      & 1040/26      & 520/13       & 120/3        \\
			Testing dataset9 with   single AR & 2120/53      & 1280/32      & 680/17       & 160/4        \\
			Testing dataset5 with   single AR & 2000/50      & 1200/30      & 680/17       & 120/3        \\
			Testing dataset6 with   single AR & 2440/61      & 1440/36      & 640/16       & 160/4        \\
			Testing dataset7 with   single AR & 2320/58      & 1320/33      & 760/19       & 80/2         \\
			Testing dataset8 with   single AR & 2400/60      & 1040/26      & 520/13       & 120/3        \\
			Testing dataset9 with   single AR & 2120/53      & 1280/32      & 680/17       & 160/4        \\ \hline
		\end{tabular}
	}
	\vspace{0.2cm}
	\raggedright
	\small
	\footnotesize{NOTE—"Mixed" means that the dataset consists of physical feature parameters from magnetograms containing multiple ARs and single AR, while "single" consists only of physical feature parameters from magnetograms containing a single AR.}
\end{table*}

\subsection{Comparison datasets} \label{subsec:Compa data}

One of the most important criteria for evaluating an operational forecasting system is the long-term accuracy of its daily predictions. To this end, we collect daily SHARP magnetograms from February 15, 2022, to June 2, 2024, for physical feature extraction and create the second type of dataset for model testing. Unlike the labeling process in Section \ref{subsec:Ten CV data}, in the process of the second type of dataset, regardless of whether the same AR appears on different days or produces multiple flares within a single day, we retain all physical feature parameters. If multiple levels of flares occur in the same AR on the same day, we only use the highest flare level of that day as the category for the AR. During the daily data acquisition process, the system first checks whether the AR contains 24 hr of magnetogram samples prior to the prediction date. If the data are insufficient, the system supplements the missing samples starting from 22:00 UT on the second day before the prediction date. If the data still do not meet the 24-hour requirement after supplementation, the AR is discarded and excluded from the second type of dataset. Similarly, we apply the z-score normalization method to this dataset, using the mean and standard deviation obtained in Section~\ref{subsec:Ten CV data}.
To enable a direct comparison with NASA/CCMC and SolarFlareNet, we design a daily mode, which serves as an evaluation method. In this mode, we align our system with existing operational forecasting systems by comparing their predictions issued at 00:00 UT for the same ARs, on the same prediction date, and over the same 24-hour forecasting window.
Our system generates predictions at 00:00 UT each day, predicting whether each AR will produce an $\geq$M–class flare within the next 24 hr. NASA/CCMC (\url{https://ccmc.gsfc.nasa.gov/scoreboards/flare/}) and SolarFlareNet (\url{https://nature.njit.edu/solardb/index.html}) also release their operational predictions at 00:00 UT daily. We retrieve their records from February 15, 2022, to June 2, 2024, and match them with the second type of dataset in daily mode.
Finally, we obtain the original testing dataset in daily mode, denoted as dataset with mixed ARs in daily mode. Similarly, for the dataset with mixed ARs in daily mode, we retain physical parameters in magnetograms containing only single AR to create a filtered dataset, referred to as dataset with single AR in daily mode. Tables \ref{tab5} and \ref{tab6} present the number distribution of samples and ARs in the dataset with single/mixed ARs in daily mode used for comparison with NASA/CCMC and SolarFlareNet, respectively.

\begin{table*}[htbp]
	\centering
	\renewcommand{\arraystretch}{1.5}
	\caption{{The number distribution of samples and ARs in the dataset with single/mixed ARs in daily mode used for comparison with NASA/CCMC.}}
	\makebox[\textwidth][c]{%
		\resizebox{1.2\textwidth}{!}{%
			\begin{tabular}{ccccc}
				\hline
				Testing datasets & N(Sample/AR) & C(Sample/AR) & M(Sample/AR) & X(Sample/AR)  \\ \hline
				Dataset with single AR in daily mode & 5680/142 & 2720/68 & 440/11 & 80/2 \\ 
				Dataset with mixed ARs in daily mode & 6720/168 & 3480/87 & 680/17 & 120/3 \\ \hline
			\end{tabular}
		}%
	}
	\label{tab5}
\end{table*}

\begin{table*}[htbp]
	\centering
	\renewcommand{\arraystretch}{1.5}
	\caption{{The number distribution of samples and ARs in the dataset with single/mixed ARs in daily mode used for comparison with SolarFlareNet.}}
	\makebox[\textwidth][c]{%
		\resizebox{1.2\textwidth}{!}{%
			\begin{tabular}{ccccc}
				\hline
				Testing datasets & N(Sample/AR) & C(Sample/AR) & M(Sample/AR) & X(Sample/AR)  \\ \hline
				Dataset with single AR in daily mode & 23440/586 & 9520/238 & 1920/48 & 120/3 \\ 
				Dataset with mixed ARs in daily mode & 25360/634 & 13400/335 & 2960/74 & 400/10 \\ \hline
			\end{tabular}
		}%
	}
	\label{tab6}
\end{table*}

\section{Method} \label{sec:Meth}

\begin{figure*}
	\centering
	\includegraphics[width=1.0\textwidth]{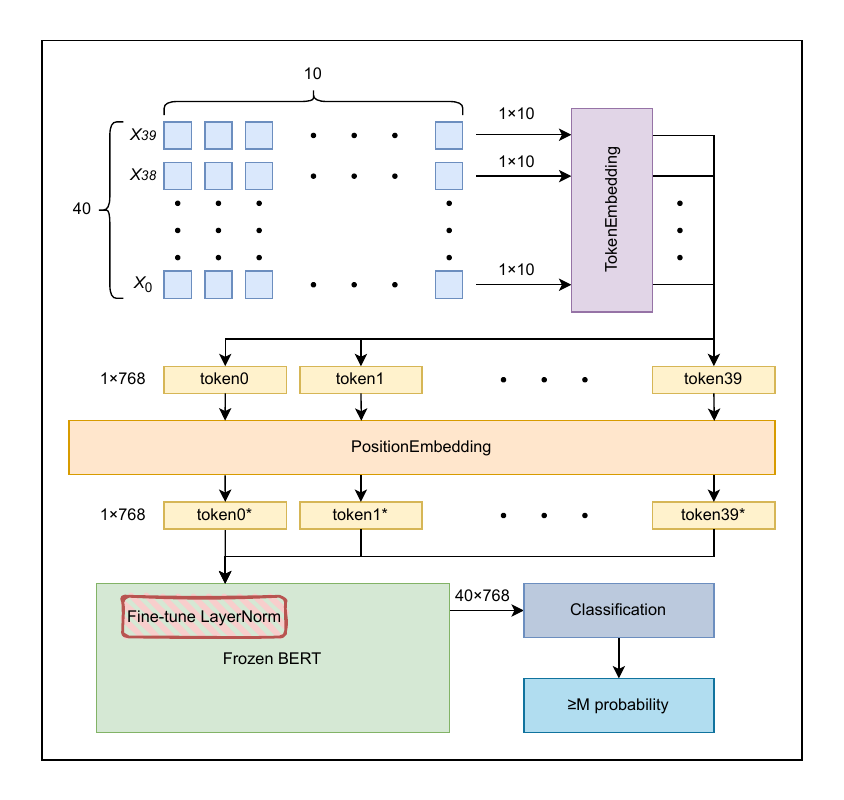}
	\caption{The model structure of LLMFlareNet.}\label{test1}
\end{figure*}

In this study, we develop an LLMFlareNet and conduct systematic ablation experiments to evaluate the effectiveness of using an LLM as a universal computation engine for solar flare forecasting.
Figure \ref{test1} illustrates the architecture of LLMFlareNet, which consists of an embedding module including TokenEmbedding layer and PositionalEmbedding layer, an LLM module, and a classification head, as described in Section \ref{subsec:fullmodel}.
In addition, we build an LSTM model as the baseline model representing traditional deep learning method and an NN model as the baseline model representing traditional machine learning method. The LSTM model consists of an LSTM module followed by a classification head, while the NN model is composed of an NN module and a classification head. The details of the baseline models are provided in Section \ref{subsec:baseline}.
Based on the LLMFlareNet model, we develop an operational flare forecasting system, with the specific scheme described in Section \ref{subsec:Sys constru}.

\subsection{LLMFlareNet} \label{subsec:fullmodel}

For clarity, we denote a sequence tensor as $[B, L, C]$, where $B$ is the batch size, $L$ is the number of time steps, and $C$ is the feature dimension. As shown in Figure~\ref{test1}, the input sequence from each AR is represented as $[X_0, X_1, \ldots, X_{38}, X_{39}]$, where $X_t \in \mathbb{R}^{[B, 1, 10]}$ denotes the 10 physical feature parameters at time step $t$ $(t \in [0, 39])$. Each sequence corresponds to 24 hr of observation data, from which one magnetogram sample is selected every 36 minutes, resulting in a total of 40 time steps.
For the 40 samples from the same AR, they all have consistent category labels. If the samples are labeled as N or C class, they are considered as negative samples; conversely, the samples labeled as M or X class are considered as positive samples.

\textbf{Embedding module.} Pre-trained LLMs are designed to capture conditional dependencies within a sequence of
discrete tokens, allowing them to extract complex contextual relationships (\citealt{vaswani2017attention,brown2020language}).
Based on this property, we segment the input time series from each AR into 40 temporal windows along time steps, each represented as a tensor of shape $[B,1,10]$. These windows are then transformed into tokens by the TokenEmbedding layer with a shape of $[B,1,768]$, corresponding to token$t$ in Figure~\ref{test1}, $t \in [0,39]$.
The TokenEmbedding layer applies a Conv1D to map each window into the embedding space suitable for LLMs, thereby converting the continuous time series into discrete tokens. Considering that the solar magnetic field exhibits clear temporal
evolution, the PositionalEmbedding layer is added after the TokenEmbedding layer to preserve the ordering information among time steps.
The resulting sequence token$t^*$ ($t^* \in [0,39]$) is then fed into the LLM module for feature extraction.

\textbf{LLM module.} \citet{zhou2023one} introduced a unified framework that applies pre-trained LLMs to time series analysis. Their theoretical and experimental results show that the self-attention mechanism can perform certain data independent operations analogous to PCA. Their finding reveals why a LLM can act as a universal computation engine. Inspired by this idea, we adopt the FPT framework (\citealt{lu2022frozen,zhou2023one}) and introduce pre-trained LLMs as a universal computation engine into the field of solar flare forecasting. After processing by the embedding module, the time series for each AR is represented as an input sequence consisting of 40 tokens, each with a shape of $[B,1,768]$. This token sequence is then passed into the LLM module, which is based on the FPT framework, for feature extraction. Under the FPT framework, we freeze all parameters except for the layer normalization modules, preserving the sequence modeling capability learned during pre-training. By fine-tuning only the layer normalization parameters, the model can better adapt to the flare forecasting task. We adopt Bidirectional Encoder Representations from Transformers (BERT; \citealt{devlin2018bert}) as the pre-trained LLM. As a pre-trained language model based on the transformer architecture, BERT differs from traditional unidirectional language models. BERT employs bidirectional context encoding, allowing it to reference both the preceding and succeeding contexts simultaneously when understanding each word. When applied to the solar flare forecasting task, BERT can leverage its bidirectional context encoding to capture the complex patterns and nonlinear relationships presented in solar activities, thereby enhancing the model ability to forecast different levels of flares. After feature extraction by the LLM module, a high-dimensional feature tensor with a shape of $[B,40,768]$ is generated. This tensor is then fed into the classification head to perform flare forecasting.

\textbf{Classification head.} 
We directly flatten the high-dimensional feature tensor produced by the LLM module to integrate information across all time steps. This design avoids additional architectural complexity and parameters, thereby highlighting the capability of BERT as a universal computation engine for feature extraction.
The flattened representation is then passed through a single linear layer that projects it into a $[B,1]$ output, followed by a Sigmoid activation to produce the probability that the AR will generate an $\geq$M-class flare within the next 24 hr.

\subsection{Baseline models} \label{subsec:baseline}

\textbf{LSTM module.} The LSTM module in this paper consists of three stacked LSTM layers, each with a hidden dimension of 512. Each layer contains multiple LSTM units, and each unit is primarily composed of a forget gate, an input gate, and an output gate \citep{van2020review}. The last hidden state from output gate is a global summary of the entire input sequence \citep{sutskever2014sequence}, containing key information and contexts from the sequence. Therefore, in this work, we use only the final hidden state of the last LSTM layer and feed it into the classification head for subsequent prediction.

\textbf{NN module.} In this study, the NN module first flattens the input time series into a one-dimensional vector, and then processes it through two fully connected layers. Each fully connected layer is followed by BatchNormalization and Dropout with a rate of 0.55. The hidden dimensions of the two fully connected layers are 128 and 32, respectively. After feature extraction in the NN module, the resulting representation is passed to the classification head for the final prediction.

\textbf{Classification head.} The baseline models adopt the same classification head design as LLMFlareNet in Section \ref{subsec:fullmodel}, only with adjustments to the input dimensionality to match the output shape of the preceding layer.

\subsection{Model parameters} \label{subsec:Mod para}

The model architectures and parameters used in Sections~\ref{subsec:fullmodel} and~\ref{subsec:baseline} are determined through multiple rounds of iterative tuning, taking into account the specific characteristics of the solar flare forecasting task. The final configurations represent the optimal settings identified through this process. The LLM module employs the bert-base-uncased model (\url{https://huggingface.co/google-bert/bert-base-uncased})
with three hidden layers. The parameter sizes of all models are summarized in Table~\ref{tab:model_params}. Although LLMFlareNet has a substantially larger total number of parameters than the baseline models, only a small fraction is trainable due to the limited fine-tuning. During training, all models use the Adam optimizer \citep{Kingma2014AdamAM} with a batch size of 16 and 50 training epochs. The initial learning rates are set to 0.00121, 0.00001, and 0.0001 for LLMFlareNet, LSTM, and NN, respectively, based on their convergence characteristics. Furthermore, we apply a learning rate scheduler that decays by a factor of 0.1 every 10 epochs to facilitate more stable convergence.

\begin{table}[htbp]
	\centering
	\caption{Total and trainable parameter sizes of LLMFlareNet and baseline models (LSTM and NN).}
	\begin{tabular}{lrr}
		\hline
		\textbf{Models} & \textbf{Total Parameters} & \textbf{Trainable Parameters} \\
		\hline
		LLMFlareNet & 45.73M & 52.23K \\
		LSTM        & 5.30M  & 5.30M  \\
		NN          & 27.99K & 27.99K \\
		\hline
	\end{tabular}
	\label{tab:model_params}
\end{table}

During the model training process, we employ the weighted binary cross-entropy loss as the loss function. To address the issues of class imbalance, the loss function incorporates class weights, thereby enhancing the focus of the models on minority classes. The specific formula is as follows:

\begin{equation}
	\text{Loss} = -\sum_{i=1}^B \left[w_1 \cdot y_{i} \cdot \log(p_{i}) + w_0 \cdot (1 - y_{i}) \cdot \log(1 - p_{i}) \right],
\end{equation}
where, Loss represents the loss value. $B$ is the total number of samples in a batch. \( y_i \) is the true label of the sample, with a value of 1 if it belongs to the positive class and 0 otherwise. \( p_i \) is the probability predicted by the model for the sample being in the positive class. \( w_1 \) and \( w_0 \) are the weights for the positive and negative classes, respectively. The specific formula is as follows:

\begin{equation}
	\label{eq:wi}
	w_i = \frac{N_{sample}}{N_{classes} \times N_{count_i}} ( i =0,1),
\end{equation}
where, \( w_i \) is the weight for class \( i \), where \( i \) takes values of 0 or 1. \( N_{count_i} \)  is the number of training samples for class \( i \), \( N_{classes} \) is the total number of classes, and \( N_{sample} \) is the total number of training samples.

\subsection{System construction\label{subsec:Sys constru}}

\begin{figure*}
	\centering
	\includegraphics[width=1.1\linewidth]{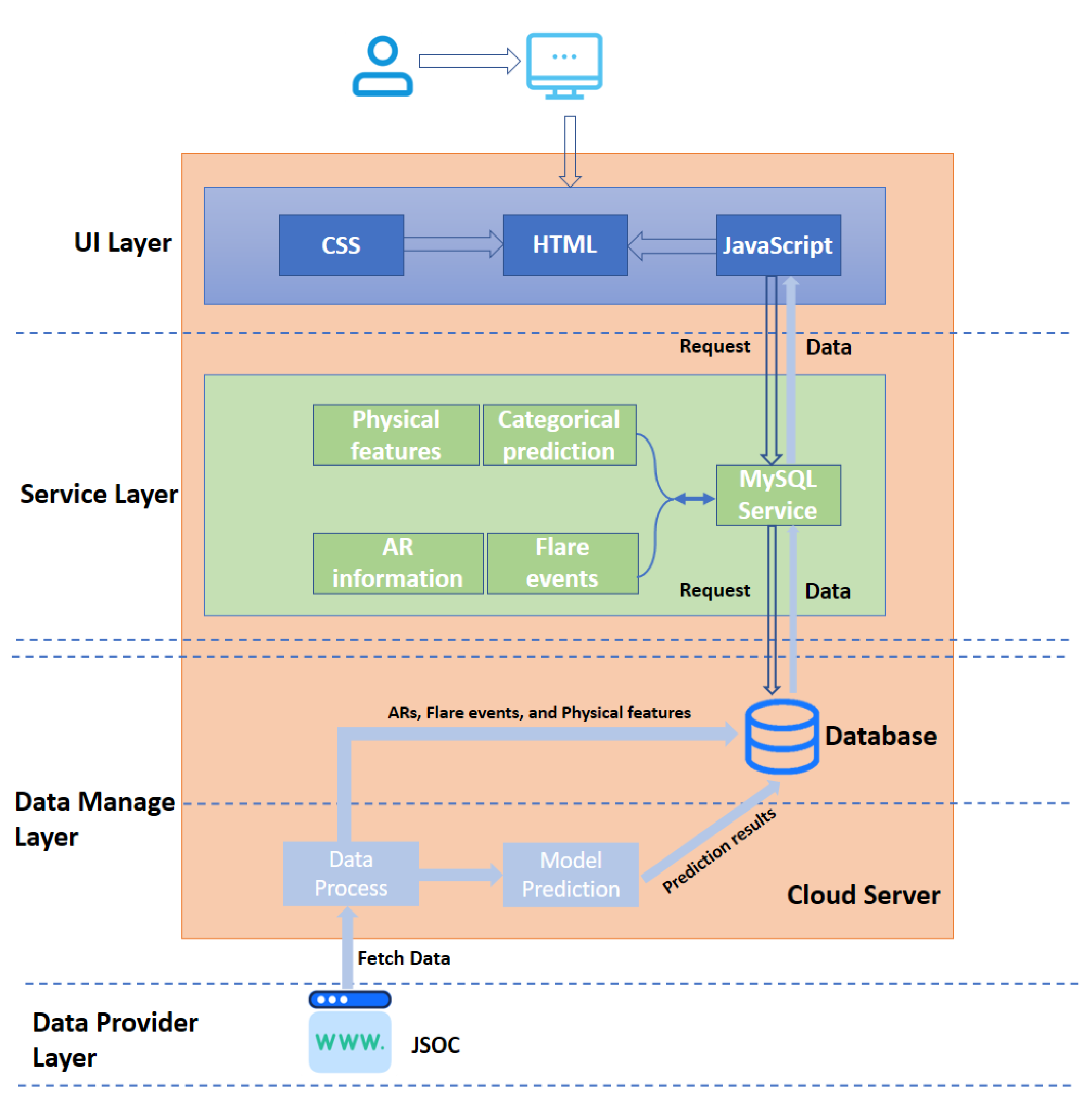}
	\caption{The architecture diagram of AR flare operational forecasting system based on the B/S architecture.}\label{figure2}
\end{figure*}

Figure \ref{figure2} illustrates the architecture diagram of our AR flare operational forecasting system based on the Browser/Server (B/S) architecture. The system primarily consists of the User Interface (UI) layer, service layer, data management layer, and data provider layer. At 00:00 UT each day, the data management layer first retrieves the daily AR information and 10 physical feature parameters from the Joint Science Operations Center (JSOC) in data provider layer. For the obtained 10 physical feature parameters, we apply the z-score normalization method to normalize the raw data, using the mean and standard deviation obtained in Section~\ref{subsec:Ten CV data} as normalization parameters. This process forms the daily AR testing data with 40 time steps.
Then, on one hand, the data management layer stores the retrieved AR information and 10 physical feature parameters into the database to facilitate the system to display historical data. On the other hand, the data management layer loads the model and performs categorical forecasting within 24 hr on the daily testing data. The system obtains the forecasting probabilities and forecasting categories, and stores the forecasting results in the database.

When users access our website of the operational forecasting system (\url{http://www.justspaceweather.cn}), the UI layer responds to different user requests, such as categorical prediction and AR information, by invoking the service layer to return the corresponding data to the UI layer. The UI layer then uses JavaScript to load the data into HTML and present it to the user.

\section{Results} \label{sec:Res}

We treat solar flare prediction in this paper as a binary classification task. On one hand, samples correctly classified as positive are defined as True Positives (TP), while samples correctly classified as negative are defined as True Negatives (TN). On the other hand, samples incorrectly predicted as positive are defined as False Positives (FP), and samples incorrectly predicted as negative are defined as False Negatives (FN). These four quantities constitute a confusion matrix. Based on the confusion matrix, we calculate multiple categorical  forecasting performance metrics, including Recall, Precision, Accuracy, Heidke Skill Score \citep{heidke1926berechnung}, True Skill Statistics (TSS; \citealt{bibTSS}), False Alarm Rate (FAR), and False Positive Rate (FPR). The TSS score varies between -1 and 1, with a score of 1 being the highest case. Similarly, the HSS score ranges from $-\infty$ to 1, with 1 signifying the optimal score. The Recall, Precision, and Accuracy scores all range from 0 to 1, with 1 representing the best score. The FAR and FPR scores range from 0 to 1, but in this case, a score of 0 is considered the best. Since TSS is not affected by class imbalance \citep{bloomfield2012toward}, we mainly use the TSS score to evaluate the categorical forecasting performance of the model. The specific formulas are as follows:

\begin{equation}
	\text{Recall} = \frac{\text{TP}}{\text{TP} + \text{FN}},
\end{equation}

\begin{equation}
	\text{Precision} = \frac{\text{TP}}{\text{TP} + \text{FP}}, 
\end{equation}

\begin{equation}
	\text{Accuracy} = \frac{\text{TP} + \text{TN}}{\text{TP} + \text{FP} + \text{TN} + \text{FN}}, 
\end{equation}

\begin{equation}
	\text{HSS} = \frac{2 (\text{TP} \times \text{TN} - \text{FP} \times \text{FN})}
	{(\text{TP} + \text{FN})(\text{FN} + \text{TN}) + (\text{TP} + \text{FP})(\text{FP} + \text{TN})}, 
\end{equation}

\begin{equation}
	\text{TSS} = \frac{\text{TP}}{\text{TP} + \text{FN}} - \frac{\text{FP}}{\text{TN} + \text{FP}}, 
\end{equation}

\begin{equation}
	\text{FAR} = \frac{\text{FP}}{\text{TP} + \text{FP}}, 
\end{equation}

\begin{equation}
	\text{FPR} = \frac{\text{FP}}{\text{FP} + \text{TN}}.
\end{equation}

\subsection{Model evaluation on the first type of dataset} \label{subsec:Mod eva}

All three models constructed in this study are trained, validated, and tested on the first type of dataset. During the training process for categorical forecasting, we monitor the TSS score of each model on the validation dataset in every epoch, continuously saving the model corresponding to the epoch with the highest TSS score on the validation dataset. The saved model is then used for testing on both the first and second types of datasets. Such a strategy helps effectively prevent overfitting. This training approach is consistent with that of \citet{li2024prediction2}.
Figures~\ref{fig:model1_loss}-\ref{fig:model6_loss} (a) and (b) illustrate the loss curves of the six models, including LLMFlareNet, two baseline models, and three ablation variants, on ten CV datasets during the training and validation processes, respectively. From these curves, it is evident that the models converge steadily and rapidly on the ten CV datasets during training process.

In this study, we conduct systematic ablation experiments to evaluate the contribution of the structure and the pre-trained knowledge within the LLM module to LLMFlareNet. Table~\ref{tab:ablation_model_params} presents the parameter sizes for different ablation variants. Table~\ref{tab:ablation_llmflarenet_vertical} reports the performance of different ablation variants on the testing dataset with mixed ARs for the $\geq$M-class flare prediction at a probability threshold of 0.5. We design three ablation configurations, including (1) completely removing the LLM module (denoted as w/o BERT Layer), (2) retaining the architecture and the FPT framework but randomly initializing the parameters of the LLM module (denoted as BERT with Random Parameters), and (3) replacing the LLM module with the Transformer Encoder (denoted as BERT $\rightarrow$ Transformer). For the first two ablation configurations, we use the same training settings as the full model. For the third configuration, the original learning rate fails to achieve stable convergence, so we reduce the initial learning rate to 0.00001 to ensure model convergence. According to Table~\ref{tab:ablation_llmflarenet_vertical}, we observe that the full LLMFlareNet model achieves a TSS of 0.720$\pm$0.040, outperforming the other three ablation variants.
This indicates the effectiveness of employing the BERT as a universal computation engine, since the model relies on the LLM module for feature extraction and sequence modeling capability. When the parameters of the LLM module are randomized, disrupting its pre-trained knowledge, the TSS decreases to 0.686$\pm$0.054. This result indicates that the performance of the model not only rely on the architecture, but also substantially benefits from the pre-trained knowledge in the BERT. Moreover, after replacing the LLM module with the Transformer Encoder of the same hidden dimension (768) and number of hidden layers (3), the TSS decreases to 0.680$\pm$0.056. Notably, although the Transformer Encoder employs a self-attention mechanism similar to BERT, it lacks large-scale pre-training. This result indicates that BERT, as a pre-trained model, is better suited for capturing long-range dependencies and complex nonlinear interactions across different features in non-stationary solar physics time series.
As shown in Table~\ref{tab:ablation_model_params}, "BERT $\rightarrow$ Transformer" has substantially more trainable parameters yet achieves inferior performance. This demonstrates that the superior performance of LLMFlareNet is not due to over-parameterization but to the  advantages of BERT as a pre-trained model in architectural design and pre-trained knowledge.
Overall, the full LLMFlareNet outperforms all three ablation variants, showing that both the structure and the pre-trained knowledge within the LLM module play important roles in enhancing the predictive performance.

\begin{table}[htbp]
	\centering
	\caption{Total and trainable parameter sizes of LLMFlareNet and three ablation variants.}
	\begin{tabular}{lrr}
		\hline
		\textbf{Models} & \textbf{Total Parameters} & \textbf{Trainable Parameters} \\
		\hline
		LLMFlareNet (Full) & 45.73M & 52.23K \\
		w/o BERT Layer & 41.47K  & 41.47K  \\
		BERT with Random Parameters  & 45.73M & 52.23K \\
		BERT $\rightarrow$ Transformer & 2.50M & 2.50M \\
		\hline
	\end{tabular}
	\label{tab:ablation_model_params}
\end{table}

In this study, we test three models (i.e., LLMFlareNet, LSTM, and NN) on the testing dataset with mixed/single ARs at a probability threshold of 0.5. Table \ref{tab:forecast_class_results} shows metric scores of each model for $\geq$M-class
flare categorical forecasting on the testing dataset with mixed/single ARs. 
On the testing dataset with mixed ARs, the LLMFlareNet model achieves TSS score of 0.720 that exceeds that of the other two baseline models (LSTM and NN) by 0.095 and 0.158, respectively, indicating that the LLMFlareNet model outperforms the other two baseline models. 
Similarly, the LLMFlareNet model achieves the highest TSS score of 0.799 on the testing dataset with single AR, outperforming the other two baseline models. Overall, on the first type of dataset, the LLMFlareNet model exhibits the best categorical forecasting performance. For the LLMFlareNet model, the TSS score of the model on testing dataset with single AR is 0.799, which is much better than that of the model on the testing dataset with mixed ARs. The similar results are also observed when each of the other two baseline models is compared on the testing dataset with mixed ARs and the testing dataset with single AR. In general, the categorical forecasting performance of the models on testing dataset with single AR is improved compared to that of models on the testing dataset with mixed ARs.

In summary, the ablation results verify the soundness of the model structure and the effectiveness of transferring pre-trained knowledge to solar flare forecasting. Furthermore, the LLMFlareNet model exhibits superior categorical forecasting performance on the first type of dataset compared to the other models. This advantage may arise from employing the pre-trained BERT as a universal computation engine. The self-attention mechanism in BERT leverages knowledge learned from massive and diverse data. This allows it to capture the complex patterns and long-range temporal evolution in solar activities, which could be difficult for LSTM or NN to learn from limited solar datasets. Therefore, we recommend the LLMFlareNet model to compare with other work in Section \ref{subsec:Compar fore}. In $\geq$M-class flare forecasting, compared to the testing dataset with mixed ARs, all three models show better performance on the testing dataset with single AR. This may be because the testing dataset with mixed ARs contains both single AR and multiple ARs, and multiple ARs may contain features of flares with different levels, leading to incorrect predictions and reducing forecasting performance.

\begin{table*}[ht]
	\centering
	\caption{Ablation study of LLMFlareNet on model architecture. The bold font highlights the best value in each row.}
	\label{tab:ablation_llmflarenet_vertical}
	\makebox[\textwidth][c]{%
		\begin{tabular}{lcccc}
			\hline
			Metrics & LLMFlareNet (Full) & w/o BERT Layer & BERT with Random Parameters & BERT $\rightarrow$ Transformer \\
			\hline
			Recall      & $\mathbf{0.894\pm0.046}$ & $0.886\pm0.046$ & $0.854\pm0.058$ & $0.840\pm0.077$ \\
			Precision   & $0.571\pm0.067$ & $0.517\pm0.054$ & $0.565\pm0.054$ & $\mathbf{0.580\pm0.073}$ \\
			Accuracy    & $0.839\pm0.034$ & $0.808\pm0.040$ & $0.836\pm0.029$ & $\mathbf{0.840\pm0.032}$ \\
			FAR         & $0.429\pm0.067$ & $0.482\pm0.054$ & $0.435\pm0.054$ & $\mathbf{0.420\pm0.073}$ \\
			FPR         & $0.174\pm0.049$ & $0.211\pm0.051$ & $0.169\pm0.041$ & $\mathbf{0.160\pm0.051}$ \\
			HSS         & $\mathbf{0.593\pm0.063}$ & $0.533\pm0.069$ & $0.574\pm0.057$ & $0.579\pm0.058$ \\
			TSS         & $\mathbf{0.720\pm0.040}$ & $0.674\pm0.063$ & $0.686\pm0.054$ & $0.680\pm0.056$ \\
			\hline
		\end{tabular}
	}
\end{table*}

\begin{table*}[ht]
	\centering
	\caption{Metric scores of each model for $\geq$M-class flare categorical forecasting on the testing dataset with mixed/single ARs. The bold font highlights the best value in each column.}
	\label{tab:forecast_class_results}
	\begin{tabular}{lccc}
		\hline
		Metrics & Model & Mixed & Single \\
		\hline
		\multirow{3}{*}{Recall}      & LLMFlareNet            & $0.894\pm0.046$ 			& $0.873\pm0.075$ \\
		& LSTM        & $0.826\pm0.075$ 			& $0.783\pm0.102$ \\
		& NN			 & $\mathbf{0.926\pm0.055}$ 			& $\mathbf{0.900\pm0.068}$ \\
		\hline
		\multirow{3}{*}{Precision}   & LLMFlareNet             & $\mathbf{0.571\pm0.067}$ 			& $\mathbf{0.733\pm0.084}$ \\
		& LSTM        & $0.511\pm0.042$ & $0.717\pm0.075$ \\
		& NN			 & $0.390\pm0.025$ 			& $0.523\pm0.074$ \\
		\hline
		\multirow{3}{*}{Accuracy}    & LLMFlareNet             & $\mathbf{0.839\pm0.034}$ 			& $\mathbf{0.916\pm0.021}$ \\
		& LSTM        & $0.805\pm0.028$ & $0.904\pm0.026$ \\
		& NN			 & $0.694\pm0.029$ 			& $0.830\pm0.040$ \\
		\hline
		\multirow{3}{*}{FAR}         & LLMFlareNet             & $\mathbf{0.429\pm0.067}$ 			& $\mathbf{0.267\pm0.084}$ \\
		& LSTM        & $0.489\pm0.042$ & $0.283\pm0.075$ \\
		& NN			 & $0.610\pm0.025$ 			& $0.477\pm0.074$ \\
		\hline
		\multirow{3}{*}{FPR}         & LLMFlareNet			& $\mathbf{0.174\pm0.049}$	& $0.074\pm0.034$ \\
		& LSTM		& $0.201\pm0.039$	& $\mathbf{0.068\pm0.022}$ \\
		& NN			& $0.364\pm0.037$	& $0.184\pm0.051$ \\
		\hline
		\multirow{3}{*}{HSS}         & LLMFlareNet             & $\mathbf{0.593\pm0.063}$ 			& $\mathbf{0.739\pm0.052}$ \\
		& LSTM        & $0.507\pm0.055$ 			& $0.686\pm0.087$ \\
		& NN			 & $0.371\pm0.043$ 			& $0.556\pm0.078$ \\
		\hline
		\multirow{3}{*}{TSS}         & LLMFlareNet             & $\mathbf{0.720\pm0.040}$ 			& $\mathbf{0.799\pm0.055}$ \\
		& LSTM        & $0.625\pm0.068$ 			& $0.715\pm0.103$ \\
		& NN			 & $0.562\pm0.057$ 			& $0.715\pm0.065$ \\
		\hline
	\end{tabular}
\end{table*}

\subsection{Model explainability analysis} \label{subsec:interpretability}
In this study, we employ the SHAP \citep{lundberg2017unified} method to explain the influence of ten physical features on the output probability of the LLMFlareNet model. The SHAP values reveal how each physical feature impacts the prediction of the model, providing an effective approach for explaining machine learning results. The SHAP values can be positive or negative. The positive or negative SHAP values indicate that the physical feature increases or decreases the prediction probability of the model, respectively. The absolute magnitude of the SHAP value reflects the extent of the influence of a physical feature on the flare prediction probability. By performing SHAP calculations on the 10 physical features for each AR, we can obtain the SHAP values of these physical  features at each time step across all ARs.

\begin{figure}
	\centering
	\includegraphics[width=0.68\textwidth]{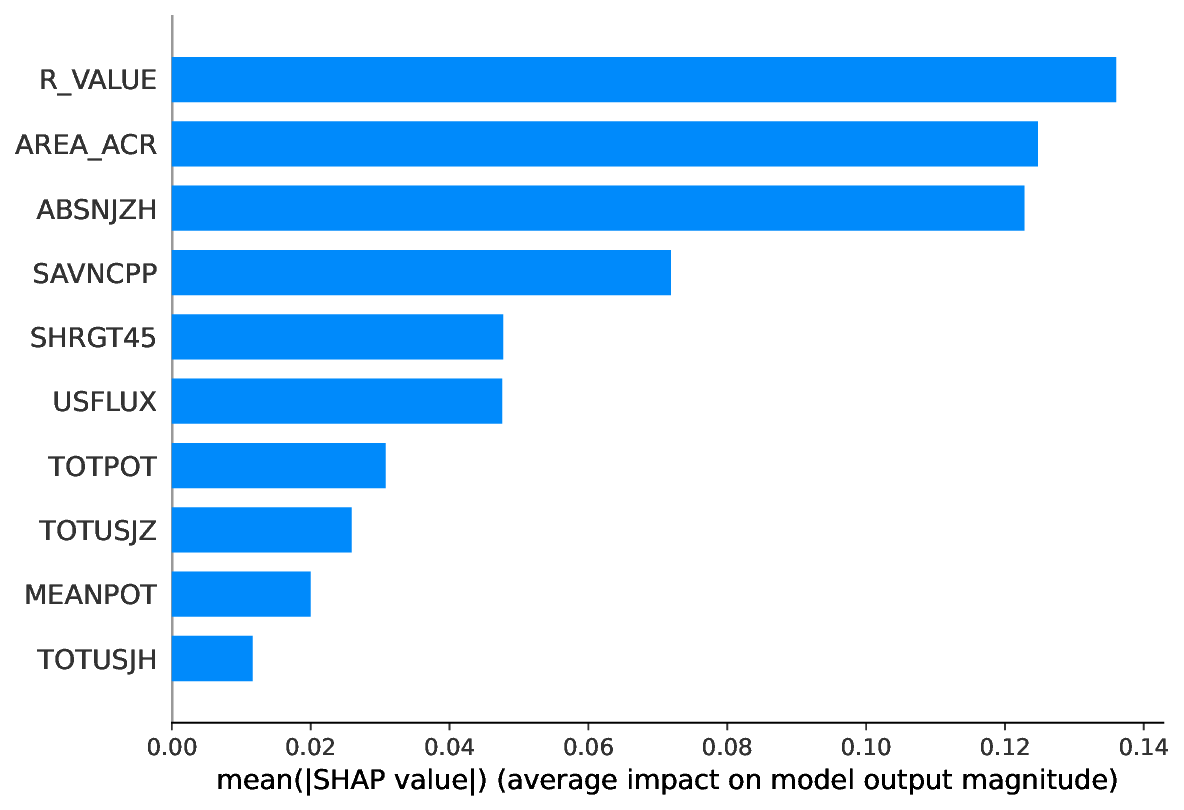}
	\caption{A bar chart illustrating the global importance of the 10 physical features for LLMFlareNet on one testing dataset from ten CV datasets. The x-axis represents the mean SHAP value and the y-axis lists the 10 physical features sorted in descending order of importance.}
	\label{fig:newfigure4}
\end{figure}

\begin{figure}
	\centering
	\includegraphics[width=0.68\textwidth]{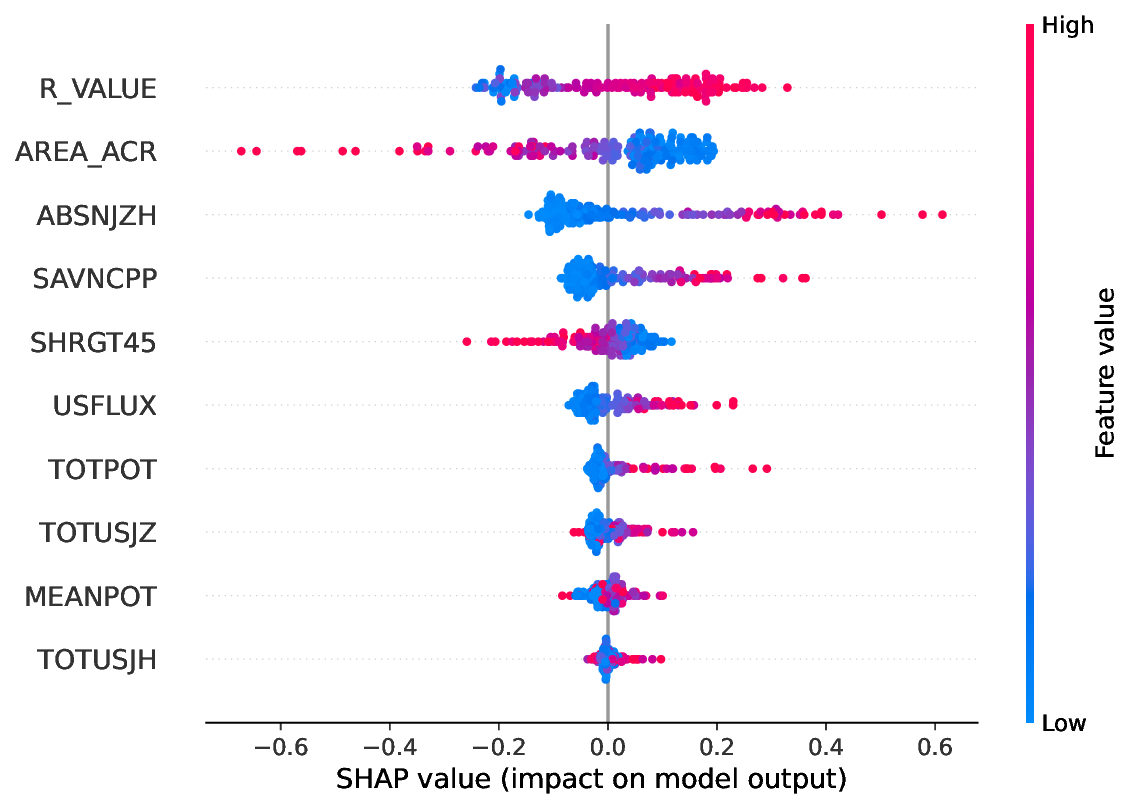}
	\caption{A beeswarm plot illustrating the impact of the ten features on LLMFlareNet for each AR. Each point corresponds to one AR. The x-axis represents the summed SHAP value of a physical feature across all time steps for each AR, with the color of the scatter points indicating the relative magnitude of the feature value.}
	\label{fig:newfigure5}
\end{figure}

Figure \ref{fig:newfigure4} presents a bar chart illustrating the global importance of the 10 physical features for LLMFlareNet on one testing dataset from ten CV datasets, with the x-axis representing the mean SHAP value and the y-axis listing the 10 physical features sorted in descending order of importance. The global mean SHAP value for a physical feature is calculated by averaging the summed SHAP values of this feature over all time steps across all ARs in the testing dataset. It is worth noting that when conducting SHAP explainability analysis of LLMFlareNet, we observe some variation in the ranking of feature importance across different CV datasets. However, a consistent finding is that the R\_VALUE emerges as the most important feature across all ten CV datasets. Therefore, the subsequent discussion focuses primarily on R\_VALUE.
As shown in Figure \ref{fig:newfigure4}, the R\_VALUE feature exhibits the highest mean SHAP value, indicating its dominant influence on the flare prediction of the LLMFlareNet model. 
Figure \ref{fig:newfigure5} displays a beeswarm plot illustrating the impact of the ten features on flare prediction for each AR. In this plot, the x-axis represents the summed SHAP value of a physical feature across all time steps for each AR, with the color of the scatter points indicating the relative magnitude of the feature value. As depicted in Figure \ref{fig:newfigure5}, larger R\_VALUE feature values correspond to higher summed SHAP values, exerting a stronger positive influence on the output probability of the LLMFlareNet model, while smaller R\_VALUE values correspond to smaller summed SHAP values, resulting in a stronger negative influence. The summed SHAP values of the R\_VALUE feature are distributed on both sides of the vertical line at SHAP value=0, clustering away from the vertical line. This indicates a significant influence on the output probability of the LLMFlareNet model. 
By averaging the summed SHAP values of each feature across all ARs, we can obtain the global mean SHAP values shown in Figure \ref{fig:newfigure4}.

\begin{figure*}
	\hspace*{-2cm} 
	\centering
	\includegraphics[width=1.25\textwidth]{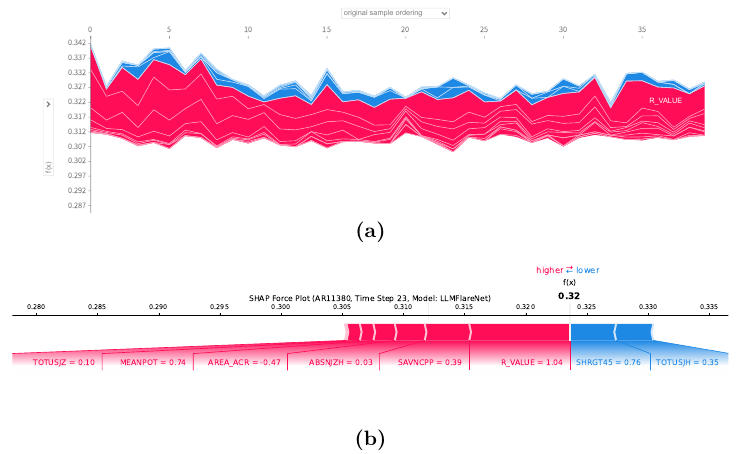}
	\caption{The force plot for the correct prediction of a positive class for AR11380.  Figure \ref{fig:newfigure6}(a) shows the force plot for AR11380 across all time steps, while Figure \ref{fig:newfigure6}(b) depicts the force plot for AR11380 at the 23rd time step. Each colored band corresponds to a physical feature. At a certain time step, the width of each band represents the SHAP value of the corresponding feature among the ten features. Red color indicates features that increase the output probability of the model, while blue color indicates features that decrease it.
	The AR11380 produced an M-class flare at 20:12 UT on December 26, 2011.}
	\label{fig:newfigure6}
\end{figure*}

\begin{figure*}
	\hspace*{-2cm} 
	\centering
	\includegraphics[width=1.25\textwidth]{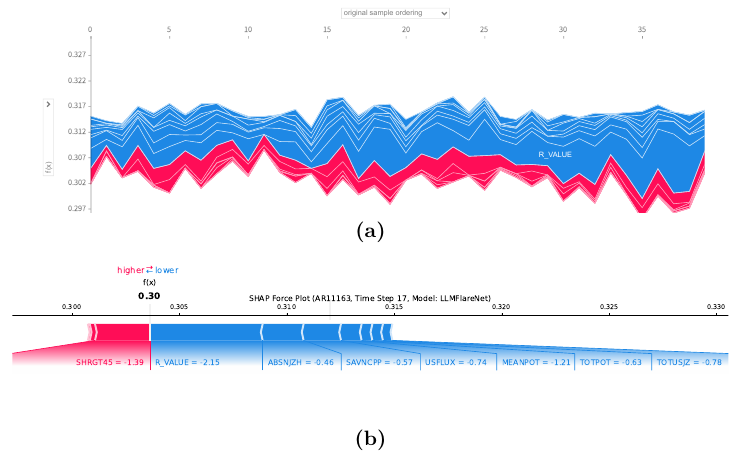}
	\caption{The force plot for the correct prediction of a negative class for AR11163.  Figure \ref{fig:newfigure7}(a) shows the force plot for AR11163 across all time steps, while Figure \ref{fig:newfigure7}(b) depicts the force plot for AR11163 at the 17th time step. 
	Each colored band corresponds to a physical feature. At a certain time step, the width of each band represents the SHAP value of the corresponding feature among the ten features. Red color indicates features that increase the output probability of the model, while blue color indicates features that decrease it.
	The AR AR11163 produced an C-class flare at 17:15 UT on March 4, 2011.}
	\label{fig:newfigure7}
\end{figure*}

To clearly clarify how the 10 physical features increase or decrease the output probability of LLMFlareNet, we randomly select one AR (e.g., AR11380) correctly predicted as positive  and one AR (e.g., AR11163) correctly predicted as negative from the testing dataset, and draw force plots for these ARs across all time steps. Additionally, we randomly select a force plot at a specific time step from above AR for detailed analysis, as shown in Figures \ref{fig:newfigure6}-\ref{fig:newfigure7}. Figure \ref{fig:newfigure6}(a) shows the force plot for AR11380 across all time steps, while Figure \ref{fig:newfigure6}(b) depicts the force plot for AR11380 at the 23th time step. In these plots, red color indicates features that increase the output probability of the model, while blue color indicates features that decrease it. As shown in Figure \ref{fig:newfigure6}(a) , the R\_VALUE feature consistently increases the output probability of the model across all time steps. In Figure \ref{fig:newfigure6}(b), the R\_VALUE feature occupies the largest red area, indicating the strongest positive impact. Together with other features in the red areas, its combined effect outweighs the negative suppression from features in blue areas, thereby aiding the model in predicting the AR as positive sample. Figure \ref{fig:newfigure7}(a) presents the force plot for AR11163 across all time steps, while Figure \ref{fig:newfigure7}(b) shows the force plot for AR11163 at the 17th time step. As shown in Figure \ref{fig:newfigure7}(a), the R\_VALUE feature continuously decreases the output probability of the model across all time steps. In  Figure \ref{fig:newfigure7}(b), the R\_VALUE feature takes over the largest blue area, indicating the most significant negative impact. Along with other features in the blue area, their cumulative effect surpasses the positive impact generated by the features in the red area, thereby facilitating the model in predicting the AR as negative sample.

To further validate the results of SHAP analysis and clarify the impact of R\_VALUE on LLMFlareNet performance, we conduct two additional ablation experiments by retraining and testing the model under different feature settings. These settings include (1) only using the R\_VALUE (denoted as Only R\_VALUE), and (2) only excluding the R\_VALUE from the ten physical features (denoted as w/o R\_VALUE). Table~\ref{tab:ablation_llmflarenet_features} presents the results of the three feature settings on the first type of testing dataset with mixed ARs at a probability threshold of 0.5. The results show that the model trained with all ten features achieves the highest TSS of 0.720, outperforming the models trained only with R\_VALUE or only without R\_VALUE. When R\_VALUE is removed from training, the TSS drops to 0.692, indicating that R\_VALUE makes a significant contribution to overall performance. Notably, the model trained only with R\_VALUE reaches the highest Recall of 0.949 among three feature settings, despite achieving the TSS of 0.668. This indicates that R\_VALUE alone still plays a strong role in predicting positive events. These results are consistent with the SHAP analysis, further confirming the critical role of R\_VALUE in the prediction performance of LLMFlareNet.

In summary, among the 10 physical features used in this study, R\_VALUE has the most crucial impact on whether LLMFlareNet can accurately forecast flare occurrence, consistent with previous findings (e.g., \citealt{liu2017predictingimportant,wei2024influence,li2024prediction2}). Previous studies primarily employed univariate feature selection algorithms and Recursive Feature Elimination (RFE) methods to investigate feature importance in flare prediction (e.g., \citealt{bobra2015solar,liu2019predicting,li2024prediction2}). Through model explainability analysis based on SHAP, we show both globally and locally how each physical feature affects the final output of the LLMFlareNet model and obtain the importance of each physical feature.
It should be emphasized that SHAP values are model-specific and may not reflect true physical causality.
In this study, R\_VALUE is identified as the most important feature on LLMFlareNet predictions. 
The R\_VALUE is defined as the total magnetic flux within a 15 Mm range around the Polarity Inversion Line (PIL) \citep{schrijver2007characteristic}, and its core physical significance lies in accurately quantifying the concentrated region of strong shear and strong gradient magnetic field required to build up magnetic free energy. 
Solar flares originate from the rapid release of free magnetic energy stored in the sheared or twisted magnetic fields of ARs through magnetic reconnection \citep{toriumi2019flare}. SHAP analysis in this study shows that R\_VALUE has the highest average SHAP value among all considered features, with a value of approximately 0.14, as shown in Figure \ref{fig:newfigure4}. Moreover, the positive contribution of R\_VALUE to flare prediction of LLMFlareNet grows when the R\_VALUE exceeds approximately 3.5, near the transition point from negative to positive SHAP values, as shown in Figure \ref{fig:newfigure5}. This phenomenon has been never reported before and it implies that the larger flux of the strong shear and gradient magnetic fields, the higher the correlation of the R\_VALUE with the generation of major flares.
The accumulation of magnetic flux in the high R\_VALUE region is essentially an energy storage process before magnetic reconnection. Its numerical growth is synchronized with the pre-flare Sigmoid structure observed by SDO/AIA \citep{biswal2024case}, confirming the physical rationality of the R\_VALUE as an indicator of magnetic free energy build-up.

\begin{table*}[ht]
	\centering
	\caption{Ablation study of LLMFlareNet on input features. The bold font highlights the best value in each row.}
	\label{tab:ablation_llmflarenet_features}
	\begin{tabular}{lccc}
		\hline
		Metrics & Full (10 Features) & Only R\_VALUE & w/o R\_VALUE \\
		\hline
		Recall      & $0.894\pm0.046$ & $\mathbf{0.949\pm0.044}$ & $0.886\pm0.049$ \\
		Precision   & $\mathbf{0.571\pm0.067}$ & $0.459\pm0.023$ & $0.538\pm0.043$ \\
		Accuracy    & $\mathbf{0.839\pm0.034}$ & $0.765\pm0.020$ & $0.822\pm0.025$ \\
		FAR         & $\mathbf{0.429\pm0.067}$ & $0.541\pm0.023$ & $0.462\pm0.043$ \\
		FPR         & $\mathbf{0.174\pm0.049}$ & $0.281\pm0.029$ & $0.194\pm0.036$ \\
		HSS         & $\mathbf{0.593\pm0.063}$ & $0.477\pm0.031$ & $0.556\pm0.046$ \\
		TSS         & $\mathbf{0.720\pm0.040}$ & $0.668\pm0.037$ & $0.692\pm0.044$ \\
		\hline
	\end{tabular}
\end{table*}

\subsection{Operational forecasting system of ARs} \label{subsec:Real fore}

Based on the recommended LLMFlareNet model and the architecture outlined in Section \ref{subsec:Sys constru}, we develop an operational forecasting system of ARs for predicting $\geq$M-class solar flares within 24 hr. we develop an operational forecasting system of ARs for solar flares within 24 hr. This system currently includes the forecasting results from the LLMFlareNet model on the second type of dataset spanning from February 15, 2022, to June 2, 2024, which are also used to compare the prediction performance of our system with that of NASA/CCMC and SolarFlareNet in Section \ref{subsec:Compar fore}.

\begin{figure*}
	\hspace*{-2cm} 
	\centering
	\includegraphics[width=1.3\textwidth]{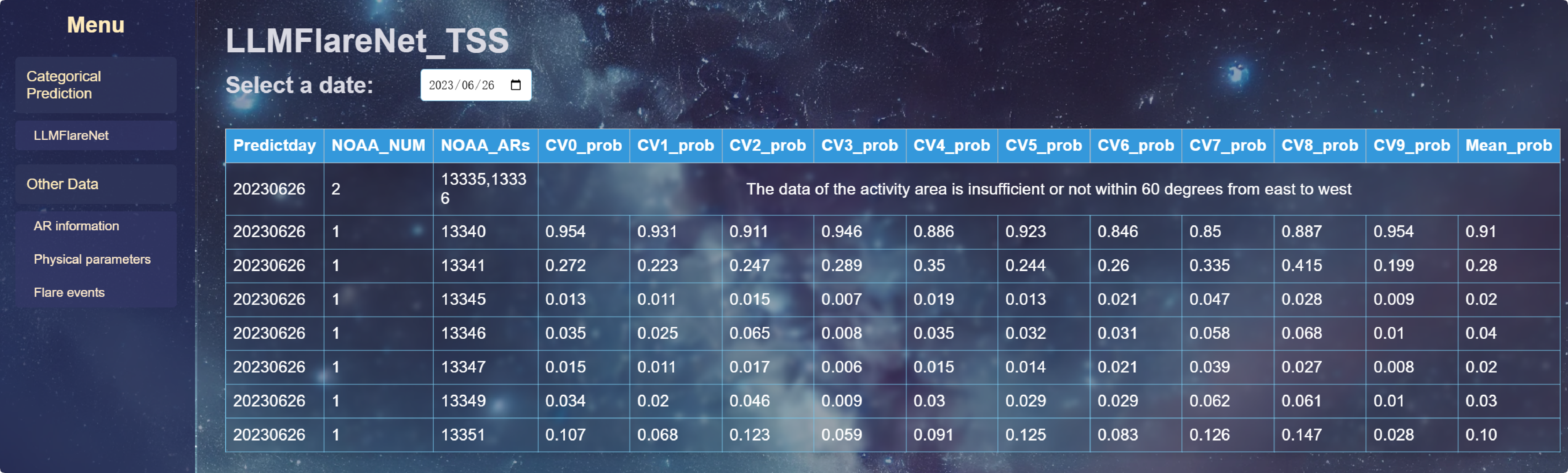}
	\caption{Graphics user interface (GUI) of categorical forecasting for LLMFlareNet model in operational forecasting system.}\label{fig4}
\end{figure*}

\begin{figure*}
	\hspace*{-2cm} 
	\centering
	\includegraphics[width=1.3\textwidth,clip, trim=5 45 235 0]{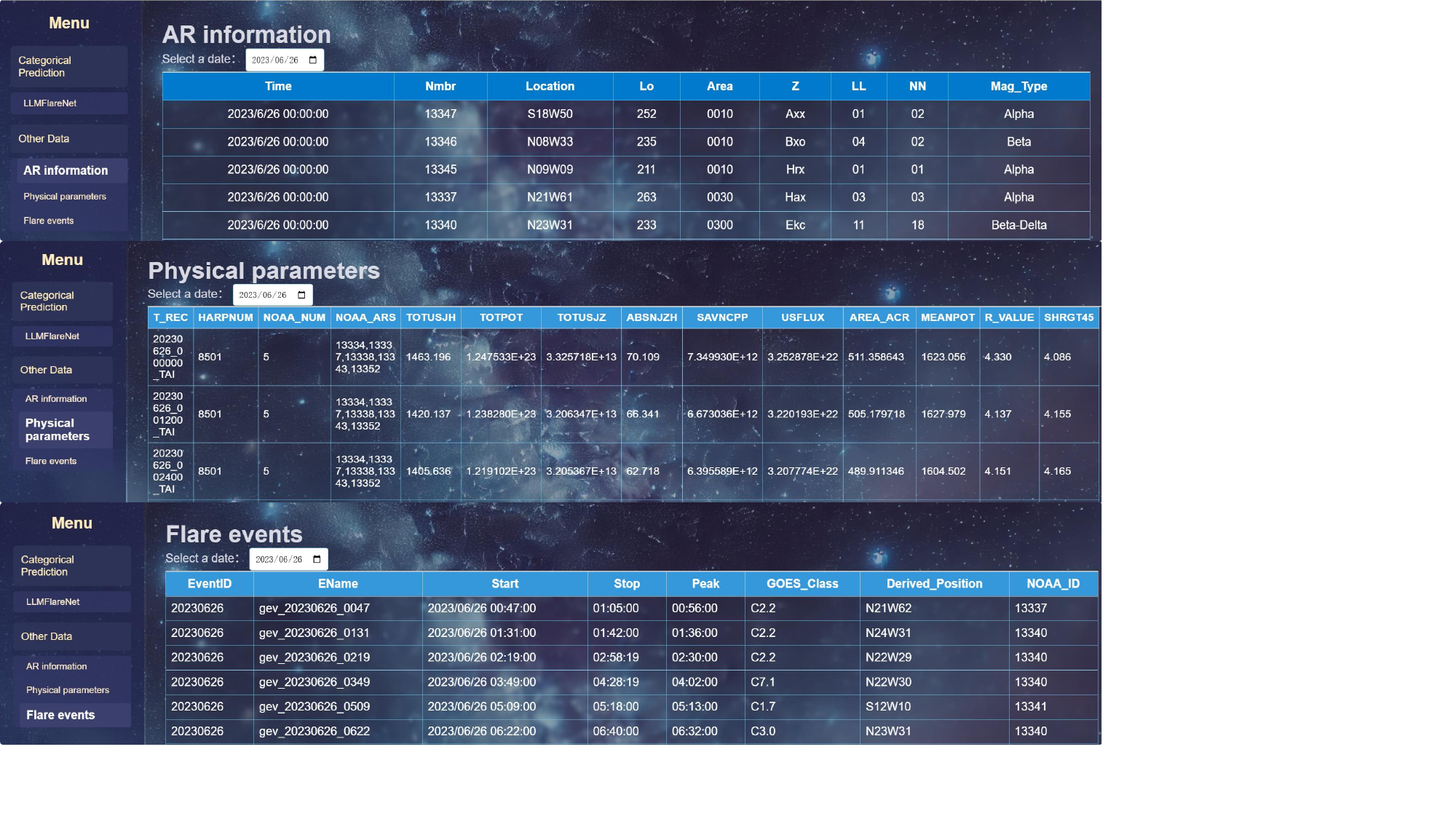}
	\caption{GUI of daily AR information, 10 physical feature parameters, and solar flare events.}\label{fig5}
\end{figure*}

In categorical forecasting, the LLMFlareNet model is trained on ten CV datasets, yielding ten trained models. The ten models then output ten probability values for each sample from the second type of dataset during prediction. We calculate the average of these ten probability values, which is used as the final forecast probability of the system for $\geq$M-class flares. 
Figure \ref{fig4} shows the graphics user interface (GUI) of categorical forecasting for LLMFlareNet model in operational forecasting system. In addition to the forecasting functionality, the system also provides daily AR information, 10 physical feature parameters, and solar flare events, as shown in Figure \ref{fig5}.

\subsection{ Comparison with available prediction} \label{subsec:Compar fore}

To objectively assess the performance of our model in forecasting $\geq$M-class
flares, we conduct tests on the second type of dataset described in Section \ref{subsec:Compa data}. Concurrently, we separately select the forecasting results from NASA/CCMC and SolarFlareNet to carry out performance comparison in daily mode
described in Section \ref{subsec:Compa data}. Additionally, since the NASA/CCMC platform integrates multiple flare prediction methods, each AR may generate multiple forecasting results daily. Considering the limited volume of prediction data from individual model at NASA/CCMC, we average the forecasting results of all models for each AR.

\begin{figure*}
	\centering
	\includegraphics[width=\textwidth]{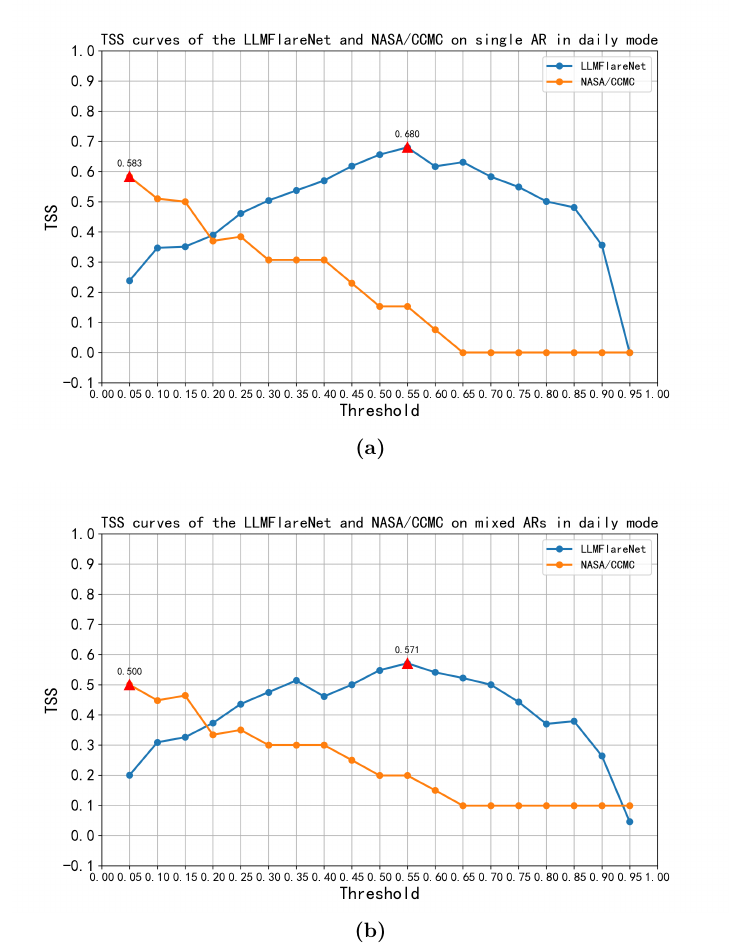}
	\caption{TSS score curves of the LLMFlareNet and NASA/CCMC with respect to probability thresholds with an increment of 0.05 on the dataset with single/mixed ARs in daily mode. Figure \ref{fig:combined4}(a) shows the TSS score curves  on the dataset with single AR and Figure \ref{fig:combined4}(b) shows the TSS score curves on the dataset with mixed ARs. Red triangles indicate the optimal TSS value on each curve.}
	\label{fig:combined4}
\end{figure*}

Based on the dataset in Table \ref{tab5} , we test the LLMFlareNet model and compare the performance of our model and NASA/CCMC. Figure \ref{fig:combined4} illustrates TSS score curves of the LLMFlareNet and NASA/CCMC with respect to probability thresholds on the dataset with single/mixed ARs in daily mode. Table \ref{tab:ccmc_alignment_vertical} shows metric scores of LLMFlareNet and NASA/CCMC for $\geq$M-class
flare categorical forecasting at the probability threshold corresponding to the optimal TSS on the dataset with single/mixed ARs in daily mode. In categorical forecasting, LLMFlareNet achieve TSS scores of 0.680/0.571 on the dataset with single/mixed ARs, which are much higher than that of NASA/CCMC at 0.583/0.500, respectively. In summary, LLMFlareNet significantly outperforms NASA/CCMC on the dataset with single/mixed ARs in daily mode.

\begin{table*}[ht]
	\centering
	\caption{Metric scores of LLMFlareNet and NASA/CCMC for $\geq$M-class flare categorical forecasting at the probability threshold corresponding to the optimal TSS on the dataset with single/mixed ARs in daily mode. The bold font highlights the best value in each column.}
	\label{tab:ccmc_alignment_vertical}
		\begin{tabular}{llcc}
			\hline
			Metrics & System & Mixed & Single \\
			\hline
			\multirow{2}{*}{Recall} & LLMFlareNet & $\mathbf{0.850}$ & $\mathbf{0.923}$ \\
			& NASA/CCMC   & $0.700$          & $0.769$          \\
			\hline
			\multirow{2}{*}{Precision} & LLMFlareNet & $0.193$          & $0.190$          \\
			& NASA/CCMC   & $\mathbf{0.215}$ & $\mathbf{0.204}$ \\
			\hline
			\multirow{2}{*}{Accuracy} & LLMFlareNet & $0.730$          & $0.766$          \\
			& NASA/CCMC   & $\mathbf{0.792}$ & $\mathbf{0.811}$ \\
			\hline
			\multirow{2}{*}{FAR} & LLMFlareNet & $0.806$          & $0.809$          \\
			& NASA/CCMC   & $\mathbf{0.784}$ & $\mathbf{0.795}$ \\
			\hline
			\multirow{2}{*}{FPR} & LLMFlareNet & $0.278$          & $0.242$          \\
			& NASA/CCMC   & $\mathbf{0.200}$ & $\mathbf{0.185}$ \\
			\hline
			\multirow{2}{*}{HSS} & LLMFlareNet & $0.222$          & $0.242$          \\
			& NASA/CCMC   & $\mathbf{0.245}$ & $\mathbf{0.253}$ \\
			\hline
			\multirow{2}{*}{TSS} & LLMFlareNet & $\mathbf{0.571}$ & $\mathbf{0.680}$ \\
			& NASA/CCMC   & $0.500$          & $0.583$          \\
			\hline
		\end{tabular}%
\end{table*}

\begin{figure*}
	\centering
	\includegraphics[width=\textwidth]{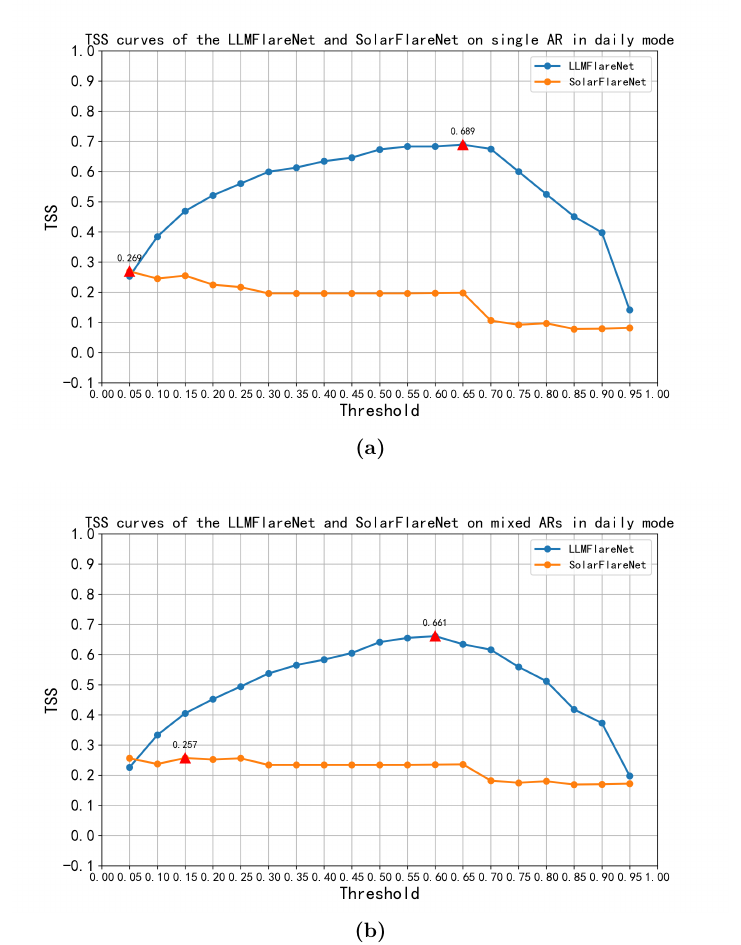}
	\caption{TSS score curves of the LLMFlareNet and SolarFlareNet with respect to probability thresholds with an increment of 0.05 on the dataset with single/mixed ARs in daily mode. Figure \ref{fig:combined5}(a) shows the TSS score curves  on the dataset with single AR and Figure \ref{fig:combined5}(b) shows the TSS score curves on the dataset with mixed ARs. Red triangles indicate the optimal TSS value on each curve.}
	\label{fig:combined5}
\end{figure*}

Based on the dataset in Table \ref{tab6}, we test the LLMFlareNet model and compare the performance of our model and SolarFlareNet. Figure \ref{fig:combined5} shows TSS score curves of the LLMFlareNet and SolarFlareNet with respect to probability thresholds on the dataset with single/mixed ARs in daily mode. Table \ref{tab:solar_alignment_vertical} shows metric scores of LLMFlareNet and SolarFlareNet for $\geq$M-class
flare categorical forecasting at the probability threshold corresponding to the optimal TSS on the dataset with single/mixed ARs in daily mode. In categorical forecasting, LLMFlareNet achieve TSS scores of 0.689/0.661 on the dataset with single/mixed ARs, which are much better than that of SolarFlareNet at 0.269/0.257, respectively. To sum up, LLMFlareNet demonstrates significantly superior forecasting performance compared to SolarFlareNet on the dataset with single/mixed ARs in daily mode.

\begin{table*}[ht]
	\centering
	\caption{Metric scores of LLMFlareNet and SolarFlareNet for $\geq$M-class flare categorical forecasting at the probability threshold corresponding to the optimal TSS on the dataset with mixed/single ARs in selected mode. The bold font highlights the best value in each column.}
	\label{tab:solar_alignment_vertical}
		\begin{tabular}{llcc}
			\hline
			Metrics & System & Mixed & Single \\
			\hline
			\multirow{2}{*}{Recall} & LLMFlareNet & $\mathbf{0.880}$ & $\mathbf{0.862}$ \\
			& SolarFlareNet & $0.440$ & $0.627$ \\
			\hline
			\multirow{2}{*}{Precision} & LLMFlareNet & $\mathbf{0.257}$ & $\mathbf{0.235}$ \\
			& SolarFlareNet & $0.172$ & $0.097$ \\
			\hline
			\multirow{2}{*}{Accuracy} & LLMFlareNet & $\mathbf{0.788}$ & $\mathbf{0.828}$ \\
			& SolarFlareNet & $0.787$ & $0.641$ \\
			\hline
			\multirow{2}{*}{FAR} & LLMFlareNet & $\mathbf{0.742}$ & $\mathbf{0.764}$ \\
			& SolarFlareNet & $0.827$ & $0.902$ \\
			\hline
			\multirow{2}{*}{FPR} & LLMFlareNet & $0.219$ & $\mathbf{0.173}$ \\
			& SolarFlareNet & $\mathbf{0.182}$ & $0.358$ \\
			\hline
			\multirow{2}{*}{HSS} & LLMFlareNet & $\mathbf{0.314}$ & $\mathbf{0.306}$ \\
			& SolarFlareNet & $0.151$ & $0.076$ \\
			\hline
			\multirow{2}{*}{TSS} & LLMFlareNet & $\mathbf{0.661}$ & $\mathbf{0.689}$ \\
			& SolarFlareNet & $0.257$ & $0.269$ \\
			\hline
		\end{tabular}%
\end{table*}

Overall, on the second type of dataset with single/mixed ARs in daily mode, the LLMFlareNet model significantly outperforms NASA/CCMC and SolarFlareNet in terms of categorical forecasting performance, respectively. This may be because the multiple advanced models within NASA/CCMC exhibit performance discrepancies, which negatively impact the overall prediction results, thereby reducing the overall forecasting performance. It is possible that our pre-trained LLM module in LLMFlareNet model could capture flare features better than the Conv1D and LSTM module in SolarFlareNet, thereby improving forecasting performance. Moreover, differences in the training strategies and training datasets used by various flare forecasting systems may also contribute to performance discrepancies, as these systems may rely on distinct training data sources, with potentially different preprocessing pipelines. In solar flare forecasting, the second type of dataset benefits from daily data collection, leading to a larger data volume and thereby rendering the forecast results more reliable. Accurate and long-term predictions of major solar flares are of paramount importance, and the daily mode in our work is more aligned with future daily prediction. By comparing the prediction results of our system with those of NASA/CCMC and SolarFlareNet in daily mode, LLMFlareNet-based system demonstrates further improved prediction performance.

\section{Conclusions and discussions} \label{sec:conclu}

In this paper, we construct two types of datasets based on SHARP data for major solar flare prediction. We develop LLMFlareNet to predict $\geq$M-class flares within 24 hr and conduct ablation experiments to verify the effectiveness of the structure and the pre-trained knowledge within the LLM module. We then compare the prediction performance of LLMFlareNet with that of baseline models (i.e., LSTM and NN). We use the model explainability method based on SHAP to explain how each physical feature influences the output probability of the LLMFlareNet model. Furthermore, to validate the SHAP analysis results, we perform additional ablation experiments on the input features.
Based on the recommended LLMFlareNet model, we develop an operational solar flare forecasting system of ARs for predicting $\geq$M-class solar flares within 24 hr. To objectively evaluate forecasting performance of the system, we compare the predictive performance of our system with that of the operational systems from NASA/CCMC and SolarFlareNet in daily mode.
This study represents the first application of large language models as a universal computation engine in the field of solar flare forecasting. It also presents the first comparison of the operational forecasting performance of the LLMFlareNet-based system with that of NASA/CCMC and SolarFlareNet in daily mode.

The main results of this paper are as follows. 
(1) On the ten CV testing dataset with mixed ARs, i.e., the first type of dataset, LLMFlareNet achieves the highest TSS of 0.720, outperforming both the baseline models and all its ablation variants. All models show higher forecasting performance on the ten CV testing dataset with single ARs than on the ten CV testing dataset with mixed ARs, with LLMFlareNet also achieving the best TSS of 0.799.
(2) Through global and local SHAP analyses, we obtain the contribution of each physical feature to the output probability of the model. R\_VALUE is found to have the greatest impact on LLMFlareNet predictions, consistent with flare magnetic reconnection theory. Moreover, additional feature ablation experiments further validate the SHAP analysis results. (3) On the dataset with single/mixed ARs in daily mode, i.e., the second type of dataset, LLMFlareNet achieves the TSS scores of {0.680/0.571 (0.689/0.661, respectively)}, significantly outperforming NASA/CCMC (SolarFlareNet, respectively). 
Overall, these results indicate that LLMs can be applied to solar flare forecasting and achieve significant improvements in both model performance and operational forecasting systems.

In this study, we adopt the comparison method that takes the daily comparison mode under the conditions of the same AR number and prediction date, which is different from previous studies in terms of performance comparison. Previous studies generally adopted one of three performance comparison methods as follows. (1) In the first comparison method, different researchers typically used different testing datasets for performance comparison       (e.g., \citealt{li2020predicting,sun2022solar}). This comparison was not based on the same data and was clearly unfair. (2) In the second comparison method, researchers compared the performance of different models based on the same dataset within their own work, without comparing them with models proposed by other researchers. This method cannot highlight the superiority or inferiority of the developed model \citep{alshammari2024transformer,abduallah2023operational}. (3) In the third comparison method, different researchers conducted performance comparisons using the same dataset \citep{grim2024solar}. This method was relatively fair, but the testing data in the dataset was also publicly available to researchers, leading to the risk of information leakage. This might allow other researchers to continuously optimize their models based on the public testing dataset, resulting in models that perform excellently on the public testing dataset but may not generalize well to other undisclosed testing dataset. In operational flare forecasting, daily AR data is constantly increasing, and future data is unknown. Therefore, optimizing models based on daily testing data is not feasible. Unlike the three comparison methods above, the performance comparison between our work and other forecasting systems (e.g., NASA/CCMC) is based on real-time observational data during the active period of solar activity under the conditions of the same AR number and prediction date. Compared with the above three methods, this approach ensures that the prediction performance comparison is more reasonable and scientific.

LLMFlareNet adopts BERT as a universal computation engine, exhibiting a general sequence modeling capability and potentially facilitating its extension to more complex multi-class flare prediction tasks. As a subsequent step, we plan to investigate LLMFlareNet for multi-class flare prediction, for instance by employing the hierarchical multiclassification scheme \citep{zheng2023multiclass}, to assess its ability to discern the subtler feature distinctions preceding flares of different classes. Such an investigation will provide stronger evidence for its general utility and sophistication. With the continuous increase of solar observation satellites, the volume and diversity of solar observational data are rapidly expanding. For instance, the Advanced Space-based Solar Observatory (ASO-S;~\citealt{gan2023advanced}) has achieved continuous multi-wavelength observations, while the upcoming Lagrange-V Solar Observatory (LAVSO;~\citealt{xihe2}, also known as “Xihe-2”) will perform stereoscopic observations from the fifth Lagrange point of the solar-terrestrial system, providing the vector magnetic fields and three-dimensional solar eruption data, further enriching available data sources. These multi-source observations will offer a more comprehensive view of the magnetic structure and dynamical evolution of ARs.
Building upon this work, we plan to develop a flare forecasting model capable of integrating multi-source data and leveraging complementary information across instruments to improve prediction performance. Leveraging this model, we will design an operational forecasting system that can handle real-time multi-source observations and provide more accurate solar flare predictions, thereby offering more reliable technical support for space weather monitoring and early warning.

\section*{Acknowledgments}
We are grateful to the anonymous reviewers whose valuable insights and feedback have significantly improved the quality of this paper. The data used here are courtesy of SDO science teams. The research was supported by the National Natural Science Foundation of China (Grant No. 12473056), the Natural Science Foundation of Jiangsu Province (Grant No. BK20241830), the B-type Strategic Priority Program of the Chinese Academy of Sciences (Grant No. XDB0560000), and the Specialized Research Fund for State
Key Laboratories.

\section*{Conflict of Interest Statement}
The authors have no conflicts of interest to disclose.

\section*{Open Research}
The data and code used in this study are available at \citep{li2025operational}.

\appendix
\section{Training Loss Curves of Different Models}
\label{appendix:loss}

\begin{figure}[h]
	\centering
	\includegraphics[width=\textwidth]{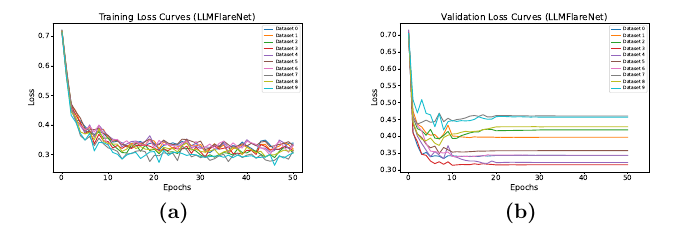}
	\caption{The training and validation loss curves of the LLMFlareNet model on ten CV datasets. Figure~\ref{fig:model1_loss}(a) represents the training loss curves, and Figure~\ref{fig:model1_loss}(b) represents the validation loss curves.}
	\label{fig:model1_loss}
\end{figure}

\begin{figure}[h]
	\centering
	\includegraphics[width=\textwidth]{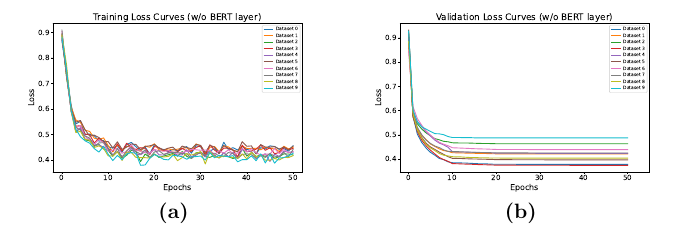}
	\caption{The training and validation loss curves of the "w/o BERT layer" model on ten CV datasets. Figure~\ref{fig:model2_loss}(a) represents the training loss curves, and Figure~\ref{fig:model2_loss}(b) represents the validation loss curves.}
	\label{fig:model2_loss}
\end{figure}

\begin{figure}[h]
	\centering
	\includegraphics[width=\textwidth]{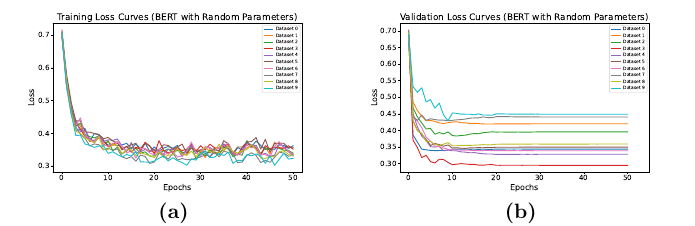}
	\caption{The training and validation loss curves of the "BERT with Random" Parameters model on ten CV datasets. Figure~\ref{fig:model3_loss}(a) represents the training loss curves, and Figure~\ref{fig:model3_loss}(b) represents the validation loss curves.}
	\label{fig:model3_loss}
\end{figure}

\begin{figure}[h]
	\centering
	\includegraphics[width=\textwidth]{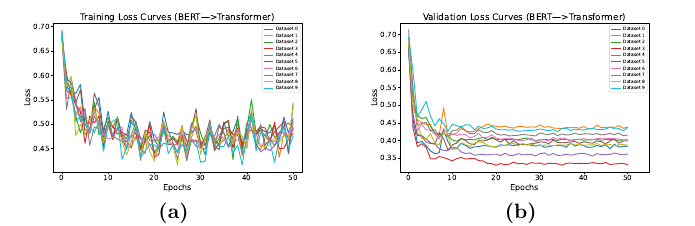}
	\caption{The training and validation loss curves of the "BERT $\rightarrow$ Transformer" model on ten CV datasets. Figure~\ref{fig:model4_loss}(a) represents the training loss curves, and Figure~\ref{fig:model4_loss}(b) represents the validation loss curves.}
	\label{fig:model4_loss}
\end{figure}

\begin{figure}[h]
	\centering
	\includegraphics[width=\textwidth]{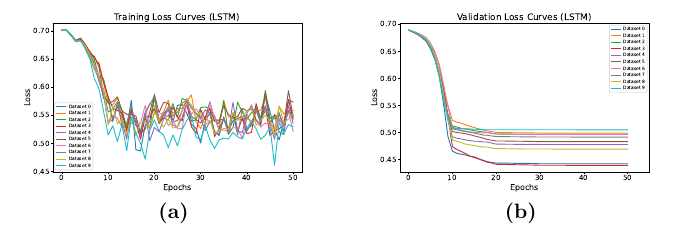}
	\caption{The training and validation loss curves of the LSTM model on ten CV datasets. Figure~\ref{fig:model5_loss}(a) represents the training loss curves, and Figure~\ref{fig:model5_loss}(b) represents the validation loss curves.}
	\label{fig:model5_loss}
\end{figure}

\begin{figure}[h]
	\centering
	\includegraphics[width=\textwidth]{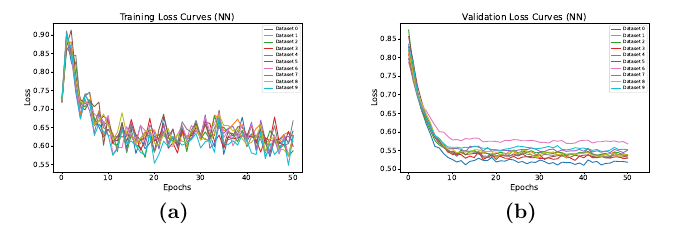}
	\caption{The training and validation loss curves of the NN model on ten CV datasets. Figure~\ref{fig:model6_loss}(a) represents the training loss curves, and Figure~\ref{fig:model6_loss}(b) represents the validation loss curves.}
	\label{fig:model6_loss}
\end{figure}

\bibliography{agujournaltemplate}

\begin{thebibliography}{}

\bibitem [\protect \citeauthoryear {%
Abduallah%
, Wang%
, Wang%
\BCBL {}\ \BBA {} Xu%
}{%
Abduallah%
\ \protect \BOthers {.}}{%
{\protect \APACyear {2023}}%
}]{%
abduallah2023operational}
\APACinsertmetastar {%
abduallah2023operational}%
\begin{APACrefauthors}%
Abduallah, Y.%
, Wang, J\BPBI T.%
, Wang, H.%
\BCBL {}\ \BBA {} Xu, Y.%
\end{APACrefauthors}%
\unskip\
\newblock
\APACrefYearMonthDay{2023}{}{}.
\newblock
{\BBOQ}\APACrefatitle {Operational prediction of solar flares using a
  transformer-based framework} {Operational prediction of solar flares using a
  transformer-based framework}.{\BBCQ}
\newblock
\APACjournalVolNumPages{Scientific reports}{13}{1}{13665}.
\newblock
\begin{APACrefDOI} \doi{10.1038/s41598-023-40884-1} \end{APACrefDOI}
\PrintBackRefs{\CurrentBib}

\bibitem [\protect \citeauthoryear {%
Al~Shalabi%
, Shaaban%
\BCBL {}\ \BBA {} Kasasbeh%
}{%
Al~Shalabi%
\ \protect \BOthers {.}}{%
{\protect \APACyear {2006}}%
}]{%
zscore}
\APACinsertmetastar {%
zscore}%
\begin{APACrefauthors}%
Al~Shalabi, L.%
, Shaaban, Z.%
\BCBL {}\ \BBA {} Kasasbeh, B.%
\end{APACrefauthors}%
\unskip\
\newblock
\APACrefYearMonthDay{2006}{}{}.
\newblock
{\BBOQ}\APACrefatitle {Data mining: A preprocessing engine} {Data mining: A
  preprocessing engine}.{\BBCQ}
\newblock
\APACjournalVolNumPages{Journal of Computer Science}{2}{9}{735--739}.
\newblock
\begin{APACrefDOI} \doi{10.3844/jcssp.2006.735.739} \end{APACrefDOI}
\PrintBackRefs{\CurrentBib}

\bibitem [\protect \citeauthoryear {%
Alshammari%
, Hamdi%
\BCBL {}\ \BBA {} Boubrahimi%
}{%
Alshammari%
\ \protect \BOthers {.}}{%
{\protect \APACyear {2024}}%
}]{%
alshammari2024transformer}
\APACinsertmetastar {%
alshammari2024transformer}%
\begin{APACrefauthors}%
Alshammari, K.%
, Hamdi, S\BPBI M.%
\BCBL {}\ \BBA {} Boubrahimi, S\BPBI F.%
\end{APACrefauthors}%
\unskip\
\newblock
\APACrefYearMonthDay{2024}{}{}.
\newblock
{\BBOQ}\APACrefatitle {Transformer Model for Multivariate Time Series
  Classification: A Case Study of Solar Flare Prediction} {Transformer model
  for multivariate time series classification: A case study of solar flare
  prediction}.{\BBCQ}
\newblock
\BIn{} \APACrefbtitle {International Conference on Pattern Recognition}
  {International conference on pattern recognition}\ (\BPGS\ 238--254).
\newblock
\begin{APACrefDOI} \doi{10.1007/978-3-031-78383-8_16} \end{APACrefDOI}
\PrintBackRefs{\CurrentBib}

\bibitem [\protect \citeauthoryear {%
Baker%
, Daly%
, Daglis%
, Kappenman%
\BCBL {}\ \BBA {} Panasyuk%
}{%
Baker%
\ \protect \BOthers {.}}{%
{\protect \APACyear {2004}}%
}]{%
baker2004effects}
\APACinsertmetastar {%
baker2004effects}%
\begin{APACrefauthors}%
Baker, D.%
, Daly, E.%
, Daglis, I.%
, Kappenman, J\BPBI G.%
\BCBL {}\ \BBA {} Panasyuk, M.%
\end{APACrefauthors}%
\unskip\
\newblock
\APACrefYearMonthDay{2004}{}{}.
\newblock
\APACrefbtitle {Effects of space weather on technology infrastructure.}
  {Effects of space weather on technology infrastructure.}
\newblock
\APACaddressPublisher{}{Wiley Online Library}.
\newblock
\begin{APACrefDOI} \doi{10.1029/2003SW000044} \end{APACrefDOI}
\PrintBackRefs{\CurrentBib}

\bibitem [\protect \citeauthoryear {%
Biswal%
\ \protect \BOthers {.}}{%
Biswal%
\ \protect \BOthers {.}}{%
{\protect \APACyear {2024}}%
}]{%
biswal2024case}
\APACinsertmetastar {%
biswal2024case}%
\begin{APACrefauthors}%
Biswal, S.%
, Kors{\'o}s, M\BPBI B.%
, Georgoulis, M\BPBI K.%
, Nindos, A.%
, Patsourakos, S.%
\BCBL {}\ \BBA {} Erd{\'e}lyi, R.%
\end{APACrefauthors}%
\unskip\
\newblock
\APACrefYearMonthDay{2024}{}{}.
\newblock
{\BBOQ}\APACrefatitle {Case Studies on Pre-eruptive X-class Flares using
  R-value in the Lower Solar Atmosphere} {Case studies on pre-eruptive x-class
  flares using r-value in the lower solar atmosphere}.{\BBCQ}
\newblock
\APACjournalVolNumPages{The Astrophysical Journal}{974}{2}{259}.
\newblock
\begin{APACrefDOI} \doi{10.3847/1538-4357/ad6c33} \end{APACrefDOI}
\PrintBackRefs{\CurrentBib}

\bibitem [\protect \citeauthoryear {%
Bloomfield%
, Higgins%
, McAteer%
\BCBL {}\ \BBA {} Gallagher%
}{%
Bloomfield%
\ \protect \BOthers {.}}{%
{\protect \APACyear {2012}}%
}]{%
bloomfield2012toward}
\APACinsertmetastar {%
bloomfield2012toward}%
\begin{APACrefauthors}%
Bloomfield, D\BPBI S.%
, Higgins, P\BPBI A.%
, McAteer, R\BPBI J.%
\BCBL {}\ \BBA {} Gallagher, P\BPBI T.%
\end{APACrefauthors}%
\unskip\
\newblock
\APACrefYearMonthDay{2012}{}{}.
\newblock
{\BBOQ}\APACrefatitle {Toward reliable benchmarking of solar flare forecasting
  methods} {Toward reliable benchmarking of solar flare forecasting
  methods}.{\BBCQ}
\newblock
\APACjournalVolNumPages{The Astrophysical Journal Letters}{747}{2}{L41}.
\newblock
\begin{APACrefDOI} \doi{10.1088/2041-8205/747/2/L41} \end{APACrefDOI}
\PrintBackRefs{\CurrentBib}

\bibitem [\protect \citeauthoryear {%
Bobra%
\ \BBA {} Couvidat%
}{%
Bobra%
\ \BBA {} Couvidat%
}{%
{\protect \APACyear {2015}}%
}]{%
bobra2015solar}
\APACinsertmetastar {%
bobra2015solar}%
\begin{APACrefauthors}%
Bobra, M\BPBI G.%
\BCBT {}\ \BBA {} Couvidat, S.%
\end{APACrefauthors}%
\unskip\
\newblock
\APACrefYearMonthDay{2015}{}{}.
\newblock
{\BBOQ}\APACrefatitle {Solar flare prediction using SDO/HMI vector magnetic
  field data with a machine-learning algorithm} {Solar flare prediction using
  sdo/hmi vector magnetic field data with a machine-learning algorithm}.{\BBCQ}
\newblock
\APACjournalVolNumPages{The Astrophysical Journal}{798}{2}{135}.
\newblock
\begin{APACrefDOI} \doi{10.1088/0004-637X/798/2/135} \end{APACrefDOI}
\PrintBackRefs{\CurrentBib}

\bibitem [\protect \citeauthoryear {%
Bobra%
\ \protect \BOthers {.}}{%
Bobra%
\ \protect \BOthers {.}}{%
{\protect \APACyear {2014}}%
}]{%
bobra2014helioseismic}
\APACinsertmetastar {%
bobra2014helioseismic}%
\begin{APACrefauthors}%
Bobra, M\BPBI G.%
, Sun, X.%
, Hoeksema, J\BPBI T.%
, Turmon, M.%
, Liu, Y.%
, Hayashi, K.%
\BDBL {}Leka, K.%
\end{APACrefauthors}%
\unskip\
\newblock
\APACrefYearMonthDay{2014}{}{}.
\newblock
{\BBOQ}\APACrefatitle {The Helioseismic and Magnetic Imager (HMI) vector
  magnetic field pipeline: SHARPs--space-weather HMI active region patches}
  {The helioseismic and magnetic imager (hmi) vector magnetic field pipeline:
  Sharps--space-weather hmi active region patches}.{\BBCQ}
\newblock
\APACjournalVolNumPages{Solar Physics}{289}{}{3549--3578}.
\newblock
\begin{APACrefDOI} \doi{10.1007/s11207-014-0529-3} \end{APACrefDOI}
\PrintBackRefs{\CurrentBib}

\bibitem [\protect \citeauthoryear {%
Brown%
\ \protect \BOthers {.}}{%
Brown%
\ \protect \BOthers {.}}{%
{\protect \APACyear {2020}}%
}]{%
brown2020language}
\APACinsertmetastar {%
brown2020language}%
\begin{APACrefauthors}%
Brown, T.%
, Mann, B.%
, Ryder, N.%
, Subbiah, M.%
, Kaplan, J\BPBI D.%
, Dhariwal, P.%
\BDBL {}Amodei, D.%
\end{APACrefauthors}%
\unskip\
\newblock
\APACrefYearMonthDay{2020}{}{}.
\newblock
{\BBOQ}\APACrefatitle {Language models are few-shot learners} {Language models
  are few-shot learners}.{\BBCQ}
\newblock
\APACjournalVolNumPages{Advances in neural information processing
  systems}{33}{}{1877--1901}.
\newblock
\begin{APACrefDOI} \doi{10.48550/arXiv.2005.14165} \end{APACrefDOI}
\PrintBackRefs{\CurrentBib}

\bibitem [\protect \citeauthoryear {%
Devlin%
, Chang%
, Lee%
\BCBL {}\ \BBA {} Toutanova%
}{%
Devlin%
\ \protect \BOthers {.}}{%
{\protect \APACyear {2018}}%
}]{%
devlin2018bert}
\APACinsertmetastar {%
devlin2018bert}%
\begin{APACrefauthors}%
Devlin, J.%
, Chang, M\BHBI W.%
, Lee, K.%
\BCBL {}\ \BBA {} Toutanova, K.%
\end{APACrefauthors}%
\unskip\
\newblock
\APACrefYearMonthDay{2018}{}{}.
\newblock
{\BBOQ}\APACrefatitle {Bert: Pre-training of deep bidirectional transformers
  for language understanding} {Bert: Pre-training of deep bidirectional
  transformers for language understanding}.{\BBCQ}
\newblock
\APACjournalVolNumPages{arXiv preprint arXiv:1810.04805}{}{}{}.
\newblock
\begin{APACrefDOI} \doi{10.48550/arXiv.1810.04805} \end{APACrefDOI}
\PrintBackRefs{\CurrentBib}

\bibitem [\protect \citeauthoryear {%
Fang%
\ \protect \BOthers {.}}{%
Fang%
\ \protect \BOthers {.}}{%
{\protect \APACyear {2024}}%
}]{%
xihe2}
\APACinsertmetastar {%
xihe2}%
\begin{APACrefauthors}%
Fang, C.%
, Ding, M.%
, Chen, P.%
, Li, C.%
, Cheng, X.%
, Guo, Y.%
\BDBL {}others%
\end{APACrefauthors}%
\unskip\
\newblock
\APACrefYearMonthDay{2024}{}{}.
\newblock
{\BBOQ}\APACrefatitle {Overview of the Lagrange-V Solar Observatory (LAVSO)}
  {Overview of the lagrange-v solar observatory (lavso)}.{\BBCQ}
\newblock
\APACjournalVolNumPages{AEROSPACE SHANGHAI}{41}{3}{9--16}.
\newblock
\begin{APACrefDOI} \doi{10.19328/j.cnki.2096⁃8655.2024.03.002}
  \end{APACrefDOI}
\PrintBackRefs{\CurrentBib}

\bibitem [\protect \citeauthoryear {%
Gan%
\ \protect \BOthers {.}}{%
Gan%
\ \protect \BOthers {.}}{%
{\protect \APACyear {2023}}%
}]{%
gan2023advanced}
\APACinsertmetastar {%
gan2023advanced}%
\begin{APACrefauthors}%
Gan, W.%
, Zhu, C.%
, Deng, Y.%
, Zhang, Z.%
, Chen, B.%
, Huang, Y.%
\BDBL {}others%
\end{APACrefauthors}%
\unskip\
\newblock
\APACrefYearMonthDay{2023}{}{}.
\newblock
{\BBOQ}\APACrefatitle {The advanced space-based solar observatory (ASO-S)} {The
  advanced space-based solar observatory (aso-s)}.{\BBCQ}
\newblock
\APACjournalVolNumPages{Solar Physics}{298}{5}{68}.
\newblock
\begin{APACrefDOI} \doi{doi.org/10.1007/s11207-023-02166-x} \end{APACrefDOI}
\PrintBackRefs{\CurrentBib}

\bibitem [\protect \citeauthoryear {%
Gazula%
, Herbert%
, Abduallah%
\BCBL {}\ \BBA {} Wang%
}{%
Gazula%
\ \protect \BOthers {.}}{%
{\protect \APACyear {2024}}%
}]{%
gazula2024interpretable}
\APACinsertmetastar {%
gazula2024interpretable}%
\begin{APACrefauthors}%
Gazula, V\BPBI R.%
, Herbert, K\BPBI G.%
, Abduallah, Y.%
\BCBL {}\ \BBA {} Wang, J\BPBI T.%
\end{APACrefauthors}%
\unskip\
\newblock
\APACrefYearMonthDay{2024}{}{}.
\newblock
{\BBOQ}\APACrefatitle {Interpretable Deep Learning for Solar Flare Prediction}
  {Interpretable deep learning for solar flare prediction}.{\BBCQ}
\newblock
\BIn{} \APACrefbtitle {2024 IEEE 36th International Conference on Tools with
  Artificial Intelligence (ICTAI)} {2024 ieee 36th international conference on
  tools with artificial intelligence (ictai)}\ (\BPGS\ 509--514).
\newblock
\begin{APACrefDOI} \doi{10.1109/ICTAI62512.2024.00078} \end{APACrefDOI}
\PrintBackRefs{\CurrentBib}

\bibitem [\protect \citeauthoryear {%
Grim%
\ \BBA {} Gradvohl%
}{%
Grim%
\ \BBA {} Gradvohl%
}{%
{\protect \APACyear {2024}}%
}]{%
grim2024solar}
\APACinsertmetastar {%
grim2024solar}%
\begin{APACrefauthors}%
Grim, L\BPBI F\BPBI L.%
\BCBT {}\ \BBA {} Gradvohl, A\BPBI L\BPBI S.%
\end{APACrefauthors}%
\unskip\
\newblock
\APACrefYearMonthDay{2024}{}{}.
\newblock
{\BBOQ}\APACrefatitle {Solar flare forecasting based on magnetogram sequences
  learning with multiscale vision transformers and data augmentation
  techniques} {Solar flare forecasting based on magnetogram sequences learning
  with multiscale vision transformers and data augmentation techniques}.{\BBCQ}
\newblock
\APACjournalVolNumPages{Solar Physics}{299}{3}{33}.
\newblock
\begin{APACrefDOI} \doi{10.1007/s11207-024-02276-0} \end{APACrefDOI}
\PrintBackRefs{\CurrentBib}

\bibitem [\protect \citeauthoryear {%
Guastavino%
, Marchetti%
, Benvenuto%
, Campi%
\BCBL {}\ \BBA {} Piana%
}{%
Guastavino%
\ \protect \BOthers {.}}{%
{\protect \APACyear {2022}}%
}]{%
guastavino2022implementation}
\APACinsertmetastar {%
guastavino2022implementation}%
\begin{APACrefauthors}%
Guastavino, S.%
, Marchetti, F.%
, Benvenuto, F.%
, Campi, C.%
\BCBL {}\ \BBA {} Piana, M.%
\end{APACrefauthors}%
\unskip\
\newblock
\APACrefYearMonthDay{2022}{}{}.
\newblock
{\BBOQ}\APACrefatitle {Implementation paradigm for supervised flare forecasting
  studies: A deep learning application with video data} {Implementation
  paradigm for supervised flare forecasting studies: A deep learning
  application with video data}.{\BBCQ}
\newblock
\APACjournalVolNumPages{Astronomy \& Astrophysics}{662}{}{A105}.
\newblock
\begin{APACrefDOI} \doi{10.1051/0004-6361/202243617} \end{APACrefDOI}
\PrintBackRefs{\CurrentBib}

\bibitem [\protect \citeauthoryear {%
Hanssen%
\ \BBA {} Kuipers%
}{%
Hanssen%
\ \BBA {} Kuipers%
}{%
{\protect \APACyear {1965}}%
}]{%
bibTSS}
\APACinsertmetastar {%
bibTSS}%
\begin{APACrefauthors}%
Hanssen, A\BPBI W.%
\BCBT {}\ \BBA {} Kuipers, W\BPBI J\BPBI A.%
\end{APACrefauthors}%
\unskip\
\newblock
\APACrefYearMonthDay{1965}{}{}.
\newblock
{\BBOQ}\APACrefatitle {Meded. Verh.} {Meded. verh.}{\BBCQ}
\newblock
\APACjournalVolNumPages{}{81}{2}{}.
\PrintBackRefs{\CurrentBib}

\bibitem [\protect \citeauthoryear {%
Heidke%
}{%
Heidke%
}{%
{\protect \APACyear {1926}}%
}]{%
heidke1926berechnung}
\APACinsertmetastar {%
heidke1926berechnung}%
\begin{APACrefauthors}%
Heidke, P.%
\end{APACrefauthors}%
\unskip\
\newblock
\APACrefYearMonthDay{1926}{}{}.
\newblock
{\BBOQ}\APACrefatitle {Berechnung des Erfolges und der G{\"u}te der
  Windst{\"a}rkevorhersagen im Sturmwarnungsdienst} {Berechnung des erfolges
  und der g{\"u}te der windst{\"a}rkevorhersagen im
  sturmwarnungsdienst}.{\BBCQ}
\newblock
\APACjournalVolNumPages{Geografiska Annaler}{8}{4}{301--349}.
\newblock
\begin{APACrefDOI} \doi{10.1080/20014422.1926.11881138} \end{APACrefDOI}
\PrintBackRefs{\CurrentBib}

\bibitem [\protect \citeauthoryear {%
Hesse%
, Bellaire%
\BCBL {}\ \BBA {} Robinson%
}{%
Hesse%
\ \protect \BOthers {.}}{%
{\protect \APACyear {2001}}%
}]{%
hesse2001community}
\APACinsertmetastar {%
hesse2001community}%
\begin{APACrefauthors}%
Hesse, M.%
, Bellaire, P.%
\BCBL {}\ \BBA {} Robinson, R.%
\end{APACrefauthors}%
\unskip\
\newblock
\APACrefYearMonthDay{2001}{}{}.
\newblock
{\BBOQ}\APACrefatitle {Community Coordinated Modeling Center: A new approach to
  space weather modeling} {Community coordinated modeling center: A new
  approach to space weather modeling}.{\BBCQ}
\newblock
\BIn{} \APACrefbtitle {Proceedings of the Space Weather Workshop: Looking
  Towards a European Space Weather Programme} {Proceedings of the space weather
  workshop: Looking towards a european space weather programme}\ (\BPGS\
  17--19).
\newblock
\begin{APACrefURL}
  \url{https://swe.ssa.esa.int/TECEES/spweather/workshops/SPW_W3/PROCEEDINGS_W3/CCMC.pdf}
  \end{APACrefURL}
\PrintBackRefs{\CurrentBib}

\bibitem [\protect \citeauthoryear {%
Huang%
\ \protect \BOthers {.}}{%
Huang%
\ \protect \BOthers {.}}{%
{\protect \APACyear {2018}}%
}]{%
RN6}
\APACinsertmetastar {%
RN6}%
\begin{APACrefauthors}%
Huang, X.%
, Wang, H.%
, Xu, L.%
, Liu, J.%
, Li, R.%
\BCBL {}\ \BBA {} Dai, X.%
\end{APACrefauthors}%
\unskip\
\newblock
\APACrefYearMonthDay{2018}{}{}.
\newblock
{\BBOQ}\APACrefatitle {Deep Learning Based Solar Flare Forecasting Model. I.
  Results for Line-of-sight Magnetograms} {Deep learning based solar flare
  forecasting model. i. results for line-of-sight magnetograms}{\BBCQ}\
  [Journal Article].
\newblock
\APACjournalVolNumPages{The Astrophysical Journal}{856}{1}{7}.
\newblock
\begin{APACrefDOI} \doi{10.3847/1538-4357/aaae00} \end{APACrefDOI}
\PrintBackRefs{\CurrentBib}

\bibitem [\protect \citeauthoryear {%
Kaneda%
\ \protect \BOthers {.}}{%
Kaneda%
\ \protect \BOthers {.}}{%
{\protect \APACyear {2022}}%
}]{%
kaneda2022flare}
\APACinsertmetastar {%
kaneda2022flare}%
\begin{APACrefauthors}%
Kaneda, K.%
, Wada, Y.%
, Iida, T.%
, Nishizuka, N.%
, Kubo, Y.%
\BCBL {}\ \BBA {} Sugiura, K.%
\end{APACrefauthors}%
\unskip\
\newblock
\APACrefYearMonthDay{2022}{}{}.
\newblock
{\BBOQ}\APACrefatitle {Flare transformer: solar flare prediction using
  magnetograms and sunspot physical features} {Flare transformer: solar flare
  prediction using magnetograms and sunspot physical features}.{\BBCQ}
\newblock
\BIn{} \APACrefbtitle {Proceedings of the Asian Conference on Computer Vision}
  {Proceedings of the asian conference on computer vision}\ (\BPGS\
  1488--1503).
\newblock
\begin{APACrefDOI} \doi{10.1007/978-3-031-26284-5_27} \end{APACrefDOI}
\PrintBackRefs{\CurrentBib}

\bibitem [\protect \citeauthoryear {%
Kingma%
\ \BBA {} Ba%
}{%
Kingma%
\ \BBA {} Ba%
}{%
{\protect \APACyear {2014}}%
}]{%
Kingma2014AdamAM}
\APACinsertmetastar {%
Kingma2014AdamAM}%
\begin{APACrefauthors}%
Kingma, D\BPBI P.%
\BCBT {}\ \BBA {} Ba, J.%
\end{APACrefauthors}%
\unskip\
\newblock
\APACrefYearMonthDay{2014}{}{}.
\newblock
{\BBOQ}\APACrefatitle {Adam: A Method for Stochastic Optimization} {Adam: A
  method for stochastic optimization}.{\BBCQ}
\newblock
\APACjournalVolNumPages{CoRR}{abs/1412.6980}{}{}.
\newblock
\begin{APACrefURL} \url{https://api.semanticscholar.org/CorpusID:6628106}
  \end{APACrefURL}
\PrintBackRefs{\CurrentBib}

\bibitem [\protect \citeauthoryear {%
Lee%
, Park%
\BCBL {}\ \BBA {} Moon%
}{%
Lee%
\ \protect \BOthers {.}}{%
{\protect \APACyear {2020}}%
}]{%
lee2020time}
\APACinsertmetastar {%
lee2020time}%
\begin{APACrefauthors}%
Lee, E\BHBI J.%
, Park, S\BHBI H.%
\BCBL {}\ \BBA {} Moon, Y\BHBI J.%
\end{APACrefauthors}%
\unskip\
\newblock
\APACrefYearMonthDay{2020}{}{}.
\newblock
{\BBOQ}\APACrefatitle {Time Series Analysis of Photospheric Magnetic Parameters
  of Flare-Quiet Versus Flaring Active Regions: Scaling Properties of
  Fluctuations} {Time series analysis of photospheric magnetic parameters of
  flare-quiet versus flaring active regions: Scaling properties of
  fluctuations}.{\BBCQ}
\newblock
\APACjournalVolNumPages{Solar Physics}{295}{9}{123}.
\newblock
\begin{APACrefDOI} \doi{10.1007/s11207-020-01690-4} \end{APACrefDOI}
\PrintBackRefs{\CurrentBib}

\bibitem [\protect \citeauthoryear {%
X.~Li%
, Li%
\BCBL {}\ \protect \BOthers {.}}{%
X.~Li%
, Li%
\BCBL {}\ \protect \BOthers {.}}{%
{\protect \APACyear {2025}}%
}]{%
li2024prediction2}
\APACinsertmetastar {%
li2024prediction2}%
\begin{APACrefauthors}%
Li, X.%
, Li, X.%
, Zheng, Y.%
, Li, T.%
, Yan, P.%
, Ye, H.%
\BDBL {}Huang, X.%
\end{APACrefauthors}%
\unskip\
\newblock
\APACrefYearMonthDay{2025}{}{}.
\newblock
{\BBOQ}\APACrefatitle {Prediction of Large Solar Flares Based on SHARP and
  High-energy-density Magnetic Field Parameters} {Prediction of large solar
  flares based on sharp and high-energy-density magnetic field
  parameters}.{\BBCQ}
\newblock
\APACjournalVolNumPages{The Astrophysical Journal Supplement
  Series}{276}{1}{7}.
\newblock
\begin{APACrefDOI} \doi{10.3847/1538-4365/ad8b2a} \end{APACrefDOI}
\PrintBackRefs{\CurrentBib}

\bibitem [\protect \citeauthoryear {%
X.~Li%
, Lv%
\BCBL {}\ \protect \BOthers {.}}{%
X.~Li%
, Lv%
\BCBL {}\ \protect \BOthers {.}}{%
{\protect \APACyear {2025}}%
}]{%
li2025operational}
\APACinsertmetastar {%
li2025operational}%
\begin{APACrefauthors}%
Li, X.%
, Lv, Y.%
, Wei, J.%
, Zheng, Y.%
, Li, T.%
, Wang, R.%
\BDBL {}Jin, H.%
\end{APACrefauthors}%
\unskip\
\newblock
\APACrefYearMonthDay{2025}{}{}.
\newblock
\APACrefbtitle {Operational Solar Flare Forecasting System Using an Explainable
  Large Language Model} {Operational solar flare forecasting system using an
  explainable large language model}\ [Software].
\newblock
\APACaddressPublisher{}{Zenodo}.
\newblock
\begin{APACrefURL} \url{https://doi.org/10.5281/zenodo.17866278}
  \end{APACrefURL}
\newblock
\begin{APACrefDOI} \doi{10.5281/zenodo.17866278} \end{APACrefDOI}
\PrintBackRefs{\CurrentBib}

\bibitem [\protect \citeauthoryear {%
X.~Li%
, Zheng%
, Wang%
\BCBL {}\ \BBA {} Wang%
}{%
X.~Li%
\ \protect \BOthers {.}}{%
{\protect \APACyear {2020}}%
}]{%
li2020predicting}
\APACinsertmetastar {%
li2020predicting}%
\begin{APACrefauthors}%
Li, X.%
, Zheng, Y.%
, Wang, X.%
\BCBL {}\ \BBA {} Wang, L.%
\end{APACrefauthors}%
\unskip\
\newblock
\APACrefYearMonthDay{2020}{}{}.
\newblock
{\BBOQ}\APACrefatitle {Predicting solar flares using a novel deep convolutional
  neural network} {Predicting solar flares using a novel deep convolutional
  neural network}.{\BBCQ}
\newblock
\APACjournalVolNumPages{The Astrophysical Journal}{891}{1}{10}.
\newblock
\begin{APACrefDOI} \doi{10.3847/1538-4357/ab6d04} \end{APACrefDOI}
\PrintBackRefs{\CurrentBib}

\bibitem [\protect \citeauthoryear {%
Y\BHBI Y.~Li%
\ \protect \BOthers {.}}{%
Y\BHBI Y.~Li%
\ \protect \BOthers {.}}{%
{\protect \APACyear {2025}}%
}]{%
li2025deep}
\APACinsertmetastar {%
li2025deep}%
\begin{APACrefauthors}%
Li, Y\BHBI Y.%
, Bai, Y.%
, Wang, C.%
, Qu, M.%
, Lu, Z.%
, Soria, R.%
\BCBL {}\ \BBA {} Liu, J.%
\end{APACrefauthors}%
\unskip\
\newblock
\APACrefYearMonthDay{2025}{}{}.
\newblock
{\BBOQ}\APACrefatitle {Deep Learning and Methods Based on Large Language Models
  Applied to Stellar Light Curve Classification} {Deep learning and methods
  based on large language models applied to stellar light curve
  classification}.{\BBCQ}
\newblock
\APACjournalVolNumPages{Intelligent Computing}{4}{}{0110}.
\newblock
\begin{APACrefDOI} \doi{10.34133/icomputing.0110} \end{APACrefDOI}
\PrintBackRefs{\CurrentBib}

\bibitem [\protect \citeauthoryear {%
C.~Liu%
, Deng%
, Wang%
\BCBL {}\ \BBA {} Wang%
}{%
C.~Liu%
\ \protect \BOthers {.}}{%
{\protect \APACyear {2017}}%
}]{%
liu2017predictingimportant}
\APACinsertmetastar {%
liu2017predictingimportant}%
\begin{APACrefauthors}%
Liu, C.%
, Deng, N.%
, Wang, J\BPBI T.%
\BCBL {}\ \BBA {} Wang, H.%
\end{APACrefauthors}%
\unskip\
\newblock
\APACrefYearMonthDay{2017}{}{}.
\newblock
{\BBOQ}\APACrefatitle {Predicting solar flares using SDO/HMI vector magnetic
  data products and the random forest algorithm} {Predicting solar flares using
  sdo/hmi vector magnetic data products and the random forest
  algorithm}.{\BBCQ}
\newblock
\APACjournalVolNumPages{The Astrophysical Journal}{843}{2}{104}.
\newblock
\begin{APACrefDOI} \doi{10.3847/1538-4357/aa789b} \end{APACrefDOI}
\PrintBackRefs{\CurrentBib}

\bibitem [\protect \citeauthoryear {%
H.~Liu%
, Liu%
, Wang%
\BCBL {}\ \BBA {} Wang%
}{%
H.~Liu%
\ \protect \BOthers {.}}{%
{\protect \APACyear {2019}}%
}]{%
liu2019predicting}
\APACinsertmetastar {%
liu2019predicting}%
\begin{APACrefauthors}%
Liu, H.%
, Liu, C.%
, Wang, J\BPBI T.%
\BCBL {}\ \BBA {} Wang, H.%
\end{APACrefauthors}%
\unskip\
\newblock
\APACrefYearMonthDay{2019}{}{}.
\newblock
{\BBOQ}\APACrefatitle {Predicting solar flares using a long short-term memory
  network} {Predicting solar flares using a long short-term memory
  network}.{\BBCQ}
\newblock
\APACjournalVolNumPages{The Astrophysical Journal}{877}{2}{121}.
\newblock
\begin{APACrefDOI} \doi{10.3847/1538-4357/ab1b3c} \end{APACrefDOI}
\PrintBackRefs{\CurrentBib}

\bibitem [\protect \citeauthoryear {%
Y.~Liu%
, Wu%
, Wang%
\BCBL {}\ \BBA {} Long%
}{%
Y.~Liu%
\ \protect \BOthers {.}}{%
{\protect \APACyear {2022}}%
}]{%
Liu2022Nonstationary}
\APACinsertmetastar {%
Liu2022Nonstationary}%
\begin{APACrefauthors}%
Liu, Y.%
, Wu, H.%
, Wang, J.%
\BCBL {}\ \BBA {} Long, M.%
\end{APACrefauthors}%
\unskip\
\newblock
\APACrefYearMonthDay{2022}{}{}.
\newblock
{\BBOQ}\APACrefatitle {Non-stationary Transformers: Exploring the Stationarity
  in Time Series Forecasting} {Non-stationary transformers: Exploring the
  stationarity in time series forecasting}.{\BBCQ}
\newblock
\BIn{} S.~Koyejo, S.~Mohamed, A.~Agarwal, D.~Belgrave, K.~Cho\BCBL {}\ \BBA {}
  A.~Oh\ (\BEDS), \APACrefbtitle {Advances in Neural Information Processing
  Systems} {Advances in neural information processing systems}\ (\BVOL~35,
  \BPGS\ 9881--9893).
\newblock
\APACaddressPublisher{}{Curran Associates, Inc.}
\newblock
\begin{APACrefURL}
  \url{https://proceedings.neurips.cc/paper_files/paper/2022/file/4054556fcaa934b0bf76da52cf4f92cb-Paper-Conference.pdf}
  \end{APACrefURL}
\PrintBackRefs{\CurrentBib}

\bibitem [\protect \citeauthoryear {%
Lu%
, Grover%
, Abbeel%
\BCBL {}\ \BBA {} Mordatch%
}{%
Lu%
\ \protect \BOthers {.}}{%
{\protect \APACyear {2022}}%
}]{%
lu2022frozen}
\APACinsertmetastar {%
lu2022frozen}%
\begin{APACrefauthors}%
Lu, K.%
, Grover, A.%
, Abbeel, P.%
\BCBL {}\ \BBA {} Mordatch, I.%
\end{APACrefauthors}%
\unskip\
\newblock
\APACrefYearMonthDay{2022}{}{}.
\newblock
{\BBOQ}\APACrefatitle {Frozen pretrained transformers as universal computation
  engines} {Frozen pretrained transformers as universal computation
  engines}.{\BBCQ}
\newblock
\BIn{} \APACrefbtitle {Proceedings of the AAAI conference on artificial
  intelligence} {Proceedings of the aaai conference on artificial
  intelligence}\ (\BVOL~36, \BPGS\ 7628--7636).
\newblock
\begin{APACrefDOI} \doi{10.1609/aaai.v36i7.20729} \end{APACrefDOI}
\PrintBackRefs{\CurrentBib}

\bibitem [\protect \citeauthoryear {%
Lundberg%
\ \BBA {} Lee%
}{%
Lundberg%
\ \BBA {} Lee%
}{%
{\protect \APACyear {2017}}%
}]{%
lundberg2017unified}
\APACinsertmetastar {%
lundberg2017unified}%
\begin{APACrefauthors}%
Lundberg, S\BPBI M.%
\BCBT {}\ \BBA {} Lee, S\BHBI I.%
\end{APACrefauthors}%
\unskip\
\newblock
\APACrefYearMonthDay{2017}{}{}.
\newblock
{\BBOQ}\APACrefatitle {A unified approach to interpreting model predictions} {A
  unified approach to interpreting model predictions}.{\BBCQ}
\newblock
\APACjournalVolNumPages{Advances in neural information processing
  systems}{30}{}{}.
\newblock
\begin{APACrefURL} \url{https://dl.acm.org/doi/10.5555/3295222.3295230}
  \end{APACrefURL}
\PrintBackRefs{\CurrentBib}

\bibitem [\protect \citeauthoryear {%
Nishizuka%
, Kubo%
, Sugiura%
, Den%
\BCBL {}\ \BBA {} Ishii%
}{%
Nishizuka%
\ \protect \BOthers {.}}{%
{\protect \APACyear {2021}}%
}]{%
nishizuka2021operational}
\APACinsertmetastar {%
nishizuka2021operational}%
\begin{APACrefauthors}%
Nishizuka, N.%
, Kubo, Y.%
, Sugiura, K.%
, Den, M.%
\BCBL {}\ \BBA {} Ishii, M.%
\end{APACrefauthors}%
\unskip\
\newblock
\APACrefYearMonthDay{2021}{}{}.
\newblock
{\BBOQ}\APACrefatitle {Operational solar flare prediction model using Deep
  Flare Net} {Operational solar flare prediction model using deep flare
  net}.{\BBCQ}
\newblock
\APACjournalVolNumPages{Earth, Planets and Space}{73}{}{1--12}.
\newblock
\begin{APACrefDOI} \doi{10.1186/s40623-021-01381-9} \end{APACrefDOI}
\PrintBackRefs{\CurrentBib}

\bibitem [\protect \citeauthoryear {%
Park%
\ \protect \BOthers {.}}{%
Park%
\ \protect \BOthers {.}}{%
{\protect \APACyear {2018}}%
}]{%
park2018application}
\APACinsertmetastar {%
park2018application}%
\begin{APACrefauthors}%
Park, E.%
, Moon, Y\BHBI J.%
, Shin, S.%
, Yi, K.%
, Lim, D.%
, Lee, H.%
\BCBL {}\ \BBA {} Shin, G.%
\end{APACrefauthors}%
\unskip\
\newblock
\APACrefYearMonthDay{2018}{}{}.
\newblock
{\BBOQ}\APACrefatitle {Application of the deep convolutional neural network to
  the forecast of solar flare occurrence using full-disk solar magnetograms}
  {Application of the deep convolutional neural network to the forecast of
  solar flare occurrence using full-disk solar magnetograms}.{\BBCQ}
\newblock
\APACjournalVolNumPages{The Astrophysical Journal}{869}{2}{91}.
\newblock
\begin{APACrefDOI} \doi{10.3847/1538-4357/aaed40} \end{APACrefDOI}
\PrintBackRefs{\CurrentBib}

\bibitem [\protect \citeauthoryear {%
Pelkum~Donahue%
\ \BBA {} Inceoglu%
}{%
Pelkum~Donahue%
\ \BBA {} Inceoglu%
}{%
{\protect \APACyear {2024}}%
}]{%
pelkum2024forecasting}
\APACinsertmetastar {%
pelkum2024forecasting}%
\begin{APACrefauthors}%
Pelkum~Donahue, K.%
\BCBT {}\ \BBA {} Inceoglu, F.%
\end{APACrefauthors}%
\unskip\
\newblock
\APACrefYearMonthDay{2024}{}{}.
\newblock
{\BBOQ}\APACrefatitle {Forecasting solar flares with a transformer network}
  {Forecasting solar flares with a transformer network}.{\BBCQ}
\newblock
\APACjournalVolNumPages{Frontiers in Astronomy and Space
  Sciences}{10}{}{1298609}.
\newblock
\begin{APACrefDOI} \doi{10.3389/fspas.2023.1298609} \end{APACrefDOI}
\PrintBackRefs{\CurrentBib}

\bibitem [\protect \citeauthoryear {%
Pesnell%
, Thompson%
\BCBL {}\ \BBA {} Chamberlin%
}{%
Pesnell%
\ \protect \BOthers {.}}{%
{\protect \APACyear {2012}}%
}]{%
pesnell2012solar}
\APACinsertmetastar {%
pesnell2012solar}%
\begin{APACrefauthors}%
Pesnell, W\BPBI D.%
, Thompson, B\BPBI J.%
\BCBL {}\ \BBA {} Chamberlin, P.%
\end{APACrefauthors}%
\unskip\
\newblock
\APACrefYear{2012}.
\newblock
\APACrefbtitle {The solar dynamics observatory (SDO)} {The solar dynamics
  observatory (sdo)}.
\newblock
\APACaddressPublisher{}{Springer}.
\newblock
\begin{APACrefDOI} \doi{10.1007/978-1-4614-3673-7_2} \end{APACrefDOI}
\PrintBackRefs{\CurrentBib}

\bibitem [\protect \citeauthoryear {%
Priest%
\ \BBA {} Forbes%
}{%
Priest%
\ \BBA {} Forbes%
}{%
{\protect \APACyear {2002}}%
}]{%
priest2002magnetic}
\APACinsertmetastar {%
priest2002magnetic}%
\begin{APACrefauthors}%
Priest, E\BPBI R.%
\BCBT {}\ \BBA {} Forbes, T.%
\end{APACrefauthors}%
\unskip\
\newblock
\APACrefYearMonthDay{2002}{}{}.
\newblock
{\BBOQ}\APACrefatitle {The magnetic nature of solar flares} {The magnetic
  nature of solar flares}.{\BBCQ}
\newblock
\APACjournalVolNumPages{The Astronomy and Astrophysics
  Review}{10}{4}{313--377}.
\newblock
\begin{APACrefDOI} \doi{10.1007/s001590100013} \end{APACrefDOI}
\PrintBackRefs{\CurrentBib}

\bibitem [\protect \citeauthoryear {%
Rawashdeh%
, Wang%
\BCBL {}\ \BBA {} Herbert%
}{%
Rawashdeh%
\ \protect \BOthers {.}}{%
{\protect \APACyear {2025}}%
}]{%
rawashdeh2025explainable}
\APACinsertmetastar {%
rawashdeh2025explainable}%
\begin{APACrefauthors}%
Rawashdeh, A\BPBI O.%
, Wang, J\BPBI T.%
\BCBL {}\ \BBA {} Herbert, K\BPBI G.%
\end{APACrefauthors}%
\unskip\
\newblock
\APACrefYearMonthDay{2025}{}{}.
\newblock
{\BBOQ}\APACrefatitle {Explainable Artificial Intelligence in Deep
  Learning-Based Solar Storm Predictions} {Explainable artificial intelligence
  in deep learning-based solar storm predictions}.{\BBCQ}
\newblock

\newblock
\begin{APACrefDOI} \doi{10.32473/flairs.38.1.138654} \end{APACrefDOI}
\PrintBackRefs{\CurrentBib}

\bibitem [\protect \citeauthoryear {%
Schou%
\ \protect \BOthers {.}}{%
Schou%
\ \protect \BOthers {.}}{%
{\protect \APACyear {2012}}%
}]{%
schou2012design}
\APACinsertmetastar {%
schou2012design}%
\begin{APACrefauthors}%
Schou, J.%
, Scherrer, P\BPBI H.%
, Bush, R\BPBI I.%
, Wachter, R.%
, Couvidat, S.%
, Rabello-Soares, M\BPBI C.%
\BDBL {}Tomczyk, S.%
\end{APACrefauthors}%
\unskip\
\newblock
\APACrefYearMonthDay{2012}{}{}.
\newblock
{\BBOQ}\APACrefatitle {Design and ground calibration of the Helioseismic and
  Magnetic Imager (HMI) instrument on the Solar Dynamics Observatory (SDO)}
  {Design and ground calibration of the helioseismic and magnetic imager (hmi)
  instrument on the solar dynamics observatory (sdo)}.{\BBCQ}
\newblock
\APACjournalVolNumPages{Solar Physics}{275}{}{229--259}.
\newblock
\begin{APACrefDOI} \doi{10.1007/s11207-011-9842-2} \end{APACrefDOI}
\PrintBackRefs{\CurrentBib}

\bibitem [\protect \citeauthoryear {%
{Schrijver}%
}{%
{Schrijver}%
}{%
{\protect \APACyear {2007}}%
}]{%
schrijver2007characteristic}
\APACinsertmetastar {%
schrijver2007characteristic}%
\begin{APACrefauthors}%
{Schrijver}, C\BPBI J.%
\end{APACrefauthors}%
\unskip\
\newblock
\APACrefYearMonthDay{2007}{{\APACmonth{02}}}{}.
\newblock
{\BBOQ}\APACrefatitle {{A Characteristic Magnetic Field Pattern Associated with
  All Major Solar Flares and Its Use in Flare Forecasting}} {{A Characteristic
  Magnetic Field Pattern Associated with All Major Solar Flares and Its Use in
  Flare Forecasting}}.{\BBCQ}
\begin{APACrefDOI} \doi{10.1086/511857} \end{APACrefDOI}
\PrintBackRefs{\CurrentBib}

\bibitem [\protect \citeauthoryear {%
Sun%
\ \protect \BOthers {.}}{%
Sun%
\ \protect \BOthers {.}}{%
{\protect \APACyear {2022}}%
}]{%
sun2022solar}
\APACinsertmetastar {%
sun2022solar}%
\begin{APACrefauthors}%
Sun, P.%
, Dai, W.%
, Ding, W.%
, Feng, S.%
, Cui, Y.%
, Liang, B.%
\BDBL {}Yang, Y.%
\end{APACrefauthors}%
\unskip\
\newblock
\APACrefYearMonthDay{2022}{}{}.
\newblock
{\BBOQ}\APACrefatitle {Solar flare forecast using 3D convolutional neural
  networks} {Solar flare forecast using 3d convolutional neural
  networks}.{\BBCQ}
\newblock
\APACjournalVolNumPages{The Astrophysical Journal}{941}{1}{1}.
\newblock
\begin{APACrefDOI} \doi{10.3847/1538-4357/ac9e53} \end{APACrefDOI}
\PrintBackRefs{\CurrentBib}

\bibitem [\protect \citeauthoryear {%
Sutskever%
, Vinyals%
\BCBL {}\ \BBA {} Le%
}{%
Sutskever%
\ \protect \BOthers {.}}{%
{\protect \APACyear {2014}}%
}]{%
sutskever2014sequence}
\APACinsertmetastar {%
sutskever2014sequence}%
\begin{APACrefauthors}%
Sutskever, I.%
, Vinyals, O.%
\BCBL {}\ \BBA {} Le, Q\BPBI V.%
\end{APACrefauthors}%
\unskip\
\newblock
\APACrefYearMonthDay{2014}{}{}.
\newblock
{\BBOQ}\APACrefatitle {Sequence to Sequence Learning with Neural Networks}
  {Sequence to sequence learning with neural networks}.{\BBCQ}.
\newblock
\begin{APACrefDOI} \doi{10.48550/arXiv.1409.3215} \end{APACrefDOI}
\PrintBackRefs{\CurrentBib}

\bibitem [\protect \citeauthoryear {%
Tang%
\ \protect \BOthers {.}}{%
Tang%
\ \protect \BOthers {.}}{%
{\protect \APACyear {2021}}%
}]{%
tang2021solar}
\APACinsertmetastar {%
tang2021solar}%
\begin{APACrefauthors}%
Tang, R.%
, Liao, W.%
, Chen, Z.%
, Zeng, X.%
, Wang, J\BHBI s.%
, Luo, B.%
\BDBL {}Wu, Z.%
\end{APACrefauthors}%
\unskip\
\newblock
\APACrefYearMonthDay{2021}{}{}.
\newblock
{\BBOQ}\APACrefatitle {Solar flare prediction based on the fusion of multiple
  deep-learning models} {Solar flare prediction based on the fusion of multiple
  deep-learning models}.{\BBCQ}
\newblock
\APACjournalVolNumPages{The Astrophysical Journal Supplement
  Series}{257}{2}{50}.
\newblock
\begin{APACrefDOI} \doi{10.3847/1538-4365/ac249e} \end{APACrefDOI}
\PrintBackRefs{\CurrentBib}

\bibitem [\protect \citeauthoryear {%
Toriumi%
\ \BBA {} Wang%
}{%
Toriumi%
\ \BBA {} Wang%
}{%
{\protect \APACyear {2019}}%
}]{%
toriumi2019flare}
\APACinsertmetastar {%
toriumi2019flare}%
\begin{APACrefauthors}%
Toriumi, S.%
\BCBT {}\ \BBA {} Wang, H.%
\end{APACrefauthors}%
\unskip\
\newblock
\APACrefYearMonthDay{2019}{}{}.
\newblock
{\BBOQ}\APACrefatitle {Flare-productive active regions} {Flare-productive
  active regions}.{\BBCQ}
\newblock
\APACjournalVolNumPages{Living Reviews in Solar Physics}{16}{1}{3}.
\newblock
\begin{APACrefDOI} \doi{10.1007/s41116-019-0019-7} \end{APACrefDOI}
\PrintBackRefs{\CurrentBib}

\bibitem [\protect \citeauthoryear {%
Van~Houdt%
, Mosquera%
\BCBL {}\ \BBA {} N{\'a}poles%
}{%
Van~Houdt%
\ \protect \BOthers {.}}{%
{\protect \APACyear {2020}}%
}]{%
van2020review}
\APACinsertmetastar {%
van2020review}%
\begin{APACrefauthors}%
Van~Houdt, G.%
, Mosquera, C.%
\BCBL {}\ \BBA {} N{\'a}poles, G.%
\end{APACrefauthors}%
\unskip\
\newblock
\APACrefYearMonthDay{2020}{}{}.
\newblock
{\BBOQ}\APACrefatitle {A review on the long short-term memory model} {A review
  on the long short-term memory model}.{\BBCQ}
\newblock
\APACjournalVolNumPages{Artificial Intelligence Review}{53}{8}{5929--5955}.
\newblock
\begin{APACrefDOI} \doi{10.1007/s10462-020-09838-1} \end{APACrefDOI}
\PrintBackRefs{\CurrentBib}

\bibitem [\protect \citeauthoryear {%
Vaswani%
\ \protect \BOthers {.}}{%
Vaswani%
\ \protect \BOthers {.}}{%
{\protect \APACyear {2017}}%
}]{%
vaswani2017attention}
\APACinsertmetastar {%
vaswani2017attention}%
\begin{APACrefauthors}%
Vaswani, A.%
, Shazeer, N.%
, Parmar, N.%
, Uszkoreit, J.%
, Jones, L.%
, Gomez, A\BPBI N.%
\BDBL {}Polosukhin, I.%
\end{APACrefauthors}%
\unskip\
\newblock
\APACrefYearMonthDay{2017}{}{}.
\newblock
{\BBOQ}\APACrefatitle {Attention is all you need} {Attention is all you
  need}.{\BBCQ}
\newblock
\APACjournalVolNumPages{Advances in Neural Information Processing
  Systems}{}{}{}.
\newblock
\begin{APACrefDOI} \doi{10.48550/arXiv.1706.03762} \end{APACrefDOI}
\PrintBackRefs{\CurrentBib}

\bibitem [\protect \citeauthoryear {%
Wei%
\ \protect \BOthers {.}}{%
Wei%
\ \protect \BOthers {.}}{%
{\protect \APACyear {2024}}%
}]{%
wei2024influence}
\APACinsertmetastar {%
wei2024influence}%
\begin{APACrefauthors}%
Wei, J.%
, Zheng, Y.%
, Li, X.%
, Xiang, C.%
, Yan, P.%
, Huang, X.%
\BDBL {}Wu, H.%
\end{APACrefauthors}%
\unskip\
\newblock
\APACrefYearMonthDay{2024}{}{}.
\newblock
{\BBOQ}\APACrefatitle {The influence of magnetic field parameters and time step
  on deep learning models of solar flare prediction} {The influence of magnetic
  field parameters and time step on deep learning models of solar flare
  prediction}.{\BBCQ}
\newblock
\APACjournalVolNumPages{Astrophysics and Space Science}{369}{5}{48}.
\newblock
\begin{APACrefDOI} \doi{10.1007/s10509-024-04314-6} \end{APACrefDOI}
\PrintBackRefs{\CurrentBib}

\bibitem [\protect \citeauthoryear {%
Yan%
\ \protect \BOthers {.}}{%
Yan%
\ \protect \BOthers {.}}{%
{\protect \APACyear {2024}}%
}]{%
yan2024real}
\APACinsertmetastar {%
yan2024real}%
\begin{APACrefauthors}%
Yan, P.%
, Li, X.%
, Zheng, Y.%
, Dong, L.%
, Yan, S.%
, Zhang, S.%
\BDBL {}Pan, Y.%
\end{APACrefauthors}%
\unskip\
\newblock
\APACrefYearMonthDay{2024}{}{}.
\newblock
{\BBOQ}\APACrefatitle {A real-time solar flare forecasting system with deep
  learning methods} {A real-time solar flare forecasting system with deep
  learning methods}.{\BBCQ}
\newblock
\APACjournalVolNumPages{Astrophysics and Space Science}{369}{10}{110}.
\newblock
\begin{APACrefDOI} \doi{10.1007/s10509-024-04374-8} \end{APACrefDOI}
\PrintBackRefs{\CurrentBib}

\bibitem [\protect \citeauthoryear {%
Ye%
, Liu%
, Hao%
\BCBL {}\ \BBA {} Cui%
}{%
Ye%
\ \protect \BOthers {.}}{%
{\protect \APACyear {2024}}%
}]{%
ye2024evaluating}
\APACinsertmetastar {%
ye2024evaluating}%
\begin{APACrefauthors}%
Ye, Y.%
, Liu, J.%
, Hao, Y.%
\BCBL {}\ \BBA {} Cui, J.%
\end{APACrefauthors}%
\unskip\
\newblock
\APACrefYearMonthDay{2024}{}{}.
\newblock
{\BBOQ}\APACrefatitle {Evaluating the Geoeffectiveness of Interplanetary
  Coronal Mass Ejections: Insights from a Support Vector Machine Approach with
  SHAP Value Analysis} {Evaluating the geoeffectiveness of interplanetary
  coronal mass ejections: Insights from a support vector machine approach with
  shap value analysis}.{\BBCQ}
\newblock
\APACjournalVolNumPages{The Astrophysical Journal}{972}{1}{52}.
\newblock
\begin{APACrefDOI} \doi{10.3847/1538-4357/ad61d7} \end{APACrefDOI}
\PrintBackRefs{\CurrentBib}

\bibitem [\protect \citeauthoryear {%
Zhang%
, Chowdhury%
, Gupta%
\BCBL {}\ \BBA {} Shang%
}{%
Zhang%
\ \protect \BOthers {.}}{%
{\protect \APACyear {2024}}%
}]{%
zhang2024large}
\APACinsertmetastar {%
zhang2024large}%
\begin{APACrefauthors}%
Zhang, X.%
, Chowdhury, R\BPBI R.%
, Gupta, R\BPBI K.%
\BCBL {}\ \BBA {} Shang, J.%
\end{APACrefauthors}%
\unskip\
\newblock
\APACrefYearMonthDay{2024}{}{}.
\newblock
{\BBOQ}\APACrefatitle {Large language models for time series: A survey} {Large
  language models for time series: A survey}.{\BBCQ}
\newblock
\APACjournalVolNumPages{arXiv preprint arXiv:2402.01801}{}{}{}.
\newblock
\begin{APACrefDOI} \doi{10.48550/arXiv.2402.01801} \end{APACrefDOI}
\PrintBackRefs{\CurrentBib}

\bibitem [\protect \citeauthoryear {%
Zheng%
, Li%
\BCBL {}\ \protect \BOthers {.}}{%
Zheng%
, Li%
\BCBL {}\ \protect \BOthers {.}}{%
{\protect \APACyear {2023}}%
}]{%
zheng2023multiclass}
\APACinsertmetastar {%
zheng2023multiclass}%
\begin{APACrefauthors}%
Zheng, Y.%
, Li, X.%
, Yan, S.%
, Huang, X.%
, Lou, H.%
\BCBL {}\ \BBA {} Li, Z.%
\end{APACrefauthors}%
\unskip\
\newblock
\APACrefYearMonthDay{2023}{}{}.
\newblock
{\BBOQ}\APACrefatitle {Multiclass solar flare forecasting models with different
  deep learning algorithms} {Multiclass solar flare forecasting models with
  different deep learning algorithms}.{\BBCQ}
\newblock
\APACjournalVolNumPages{Monthly Notices of the Royal Astronomical
  Society}{521}{4}{5384--5399}.
\newblock
\begin{APACrefDOI} \doi{10.1093/mnras/stad839} \end{APACrefDOI}
\PrintBackRefs{\CurrentBib}

\bibitem [\protect \citeauthoryear {%
Zheng%
, Qin%
\BCBL {}\ \protect \BOthers {.}}{%
Zheng%
, Qin%
\BCBL {}\ \protect \BOthers {.}}{%
{\protect \APACyear {2023}}%
}]{%
zheng2023comparative}
\APACinsertmetastar {%
zheng2023comparative}%
\begin{APACrefauthors}%
Zheng, Y.%
, Qin, W.%
, Li, X.%
, Ling, Y.%
, Huang, X.%
, Li, X.%
\BDBL {}Lou, H.%
\end{APACrefauthors}%
\unskip\
\newblock
\APACrefYearMonthDay{2023}{}{}.
\newblock
{\BBOQ}\APACrefatitle {Comparative analysis of machine learning models for
  solar flare prediction} {Comparative analysis of machine learning models for
  solar flare prediction}.{\BBCQ}
\newblock
\APACjournalVolNumPages{Astrophysics and Space Science}{368}{7}{53}.
\newblock
\begin{APACrefDOI} \doi{10.1007/s10509-023-04209-y} \end{APACrefDOI}
\PrintBackRefs{\CurrentBib}

\bibitem [\protect \citeauthoryear {%
Zhou%
, Niu%
, Sun%
, Jin%
\BCBL {}\ \protect \BOthers {.}}{%
Zhou%
\ \protect \BOthers {.}}{%
{\protect \APACyear {2023}}%
}]{%
zhou2023one}
\APACinsertmetastar {%
zhou2023one}%
\begin{APACrefauthors}%
Zhou, T.%
, Niu, P.%
, Sun, L.%
, Jin, R.%
\BCBL {}\ \BOthersPeriod {.}\end{APACrefauthors}%
\unskip\
\newblock
\APACrefYearMonthDay{2023}{}{}.
\newblock
{\BBOQ}\APACrefatitle {One fits all: Power general time series analysis by
  pretrained lm} {One fits all: Power general time series analysis by
  pretrained lm}.{\BBCQ}
\newblock
\APACjournalVolNumPages{Advances in neural information processing
  systems}{36}{}{43322--43355}.
\newblock
\begin{APACrefURL}
  \url{https://proceedings.neurips.cc/paper_files/paper/2023/file/86c17de05579cde52025f9984e6e2ebb-Paper-Conference.pdf}
  \end{APACrefURL}
\PrintBackRefs{\CurrentBib}

\end{thebibliography}

\end{document}